\documentstyle[12pt]{article}
\input epsf
\makeatletter

\@addtoreset{equation}{section}
\makeatother
\newcommand{\tselea}[1]{\label{#1}}
\newcommand{\tseleq}[1]{\label{#1}}

\newcommand{\tseref}[1]{\ref{#1}} 
\newcommand{\tsecite}[1]{{\small{$^{\cite {#1}}$}}} 

\newcommand{\tsebibitem}[1]{\bibitem{#1}} 

\newcommand{\vect}[1]{\bf #1}

\newcommand{\cali}[1]{\cal #1}

\begin{document}

\typeout{--- Title page start ---}

\renewcommand{\thefootnote}{\fnsymbol{footnote}}

\vskip 4cm

\begin{center}{\Large\bf Heat Bath
Particle Number Spectrum} \vskip 1.2cm {\large
P.Jizba\footnote{E-mail:
{\tt pj10006@damtp.cam.ac.uk}}}\\ {\em DAMTP, University of Cambridge,
Silver Street, Cambridge, CB3 9EW, UK} \end{center} 

\vskip 1cm
\begin{center} {\large\bf Abstract}
\end{center}

\begin{minipage}{14.92cm} {\small We calculate the number spectrum of
particles radiated during a scattering into a heat bath using the thermal
largest-time equation and the Dyson-Schwinger equation. We show how one
can systematically calculate $\frac{d\langle N(\omega) \rangle}{d\omega}$
to any order using modified real time finite-temperature diagrams. 
Our approach is demonstrated on a simple model where two scalar particles
scatter, within a photon-electron heat bath, into a pair of charged
particles and it is shown how to calculate the resulting changes in the
number spectra of the photons and electrons.} \end{minipage}

\vskip 1.5cm

\renewcommand{\thefootnote}{\fnsymbol{footnote}}
\setcounter{footnote}{0}

\typeout{--- Main Text Start ---}

\section{Introduction}

In recent years much theoretical effort has been invested in the understanding of
relativistic heavy ion collisions as these can create critical energy
densities which are large enough to produce the quark-gluon plasma (the
deconfined phase of quarks and gluons) \tsecite{LB, LW}. 

\vspace{3mm}

\noindent A natural tool for testing the quark-gluon plasma properties
would be to look for the particle number spectrum formed when a particle
decays within the plasma itself.  As the plasma created during heavy ion
collisions is, to a very good approximation, in thermodynamical
equilibrium \tsecite{LB} (somewhat like a ``microwave oven'' or a heat
bath), one can use the whole machinery of statistical physics and QFT in
order to predict the final plasma number spectrum. Such calculations,
derived from first principles, were carried out by Landshoff and Taylor
\tsecite{PVLJT}. 

\vspace{3mm}

\noindent Our aim was to find a sufficiently easy mathematical formalism
allowing us to perform mentioned calculations to any order. Because
unstable particles treated in \tsecite{PVLJT} can not naturally appear in
asymptotic states, we demonstrate our approach on a mathematically more
correct (but from practical point of view less relevant) process; namely
on the scattering of two particles inside of a heat bath. The method
presented here however, might be applied as well to a decay itself
(provided that the corresponding decay rate is much less than any of the
characteristic energies of the process). In this paper we formulate the
basic diagrammatic rules for the methodical perturbative calculus of
plasma particle number spectrum $\frac{d\langle N(\omega)
\rangle}{d\omega} $ and discuss it in the simple case of a heat bath
comprised of photons and electrons, which are for simplicity treated as
scalar particles. In Section 2 we review the basic concepts and techniques
needed from the theory of the largest-time equation (both for $T=0$ and $T
\not=0$) and the Dyson- Schwinger equation.  Rules for the cut diagrams at
finite-temperature are derived and subsequently extended to the case when
unheated fields are present. It was already pointed out in \tsecite{PL2}
that the thermal cut diagrams are virtually the Kobes-Semenoff diagrams
\tsecite{LB} in the Keldysh formalism \tsecite{K1}. This observation will
allow us to identify type 1 vertices in the real time finite-temperature
diagrams with the uncircled vertices used in the (thermal) cut diagrams,
and similarly type 2 vertices will be identified with the circled, cut
diagram vertices. As we want to restrict our attention to only some
particular final particle states, further restrictions on the possible cut
diagrams must be included. We shall study these restrictions in the last
part of Section 2. 

\vspace{3mm}

\noindent As we shall show in Section 3, the heat-bath particle number
spectrum can be conveniently expressed as a fraction. Whilst it is
possible to compute the denominator by means of the thermal cut diagrams
developed in Section 2, the calculation of the numerator requires more
care. Using the Dyson-Schwinger equation, we shall see in Section 4 that
it can be calculated through modified thermal cut diagrams. The
modification consists of the substitution in turn of each heat bath
particle propagator by an altered one. We also show that there must be
only one modification per diagram. From this we conclude that from each
individual cut diagram we get $n$ modified ones ($n$ stands for the total
number of heat-bath particle propagators in the diagram). Furthermore, in
the case when more types of the heat bath particles are present, one
might be only interested in the number spectrum of some of these. The
construction of the modified cut diagrams in such cases follow the same
procedure as in the previous situation. We find that only the propagators
affiliated to the desired fields must be altered. 

\vspace{3mm}

\noindent In Section 5 the presented approach is applied to a toy model in
which the quark-gluon plasma is simulated by {\em {scalar}} photons and
electrons and we calculate the resulting changes in the number spectrum of
the ``plasma'' particles. 

\vspace{3mm}

\noindent Finally, in Appendix A we derive, directly from the thermal Wick's
theorem, the (thermal) Dyson-Schwinger equation as well as other useful
functional identities valid at finite temperature. 

\section{Basic tools}
\subsection{Mean statistical value}

The central idea of thermal QFT is based on the fact that one can not take the
expectation value of an observable $A$ with respect to some pure state as
generally all states have non-zero probability to be populated and
consequently one must consider instead a mixture of states generally
described by the density matrix $\rho$. The mean statistical value
of $A$ is then

\begin{equation}
\langle A \rangle = Tr(\rho A),
\tseleq{mean value}
\end{equation}

\vspace{2mm}

\noindent where the trace has to be taken over a complete set of {\em
physical} states. For a statistical system in thermodynamical equilibrium
$\rho$ is given by the Gibbs canonical distribution

\begin{equation} \rho = \frac{e^{-\beta (H - \mu N)}}{Tr(e^{-\beta (H - \mu
N)})} = \frac{e^{-\beta K }}{Z},
\end{equation}

\noindent here $Z$ is the partition function, $H$ is the Hamiltonian, $N$
is the conserved charge (e.g. baryon or lepton number), $\mu$ is the
chemical potential, $K = H - \mu N$, and $\beta$ is the inverse
temperature: $\beta = 1/T$ ($k_{B}=1$).

\subsection{Largest-time equation at T=0}

An important property inherited from zero-temperature QFT is {\em the
largest-time equation} (LTE) \tsecite{PN, V, TV}. Although the following
sections will mainly hinge on the {\em thermal} LTE, it is instructive to
start first with the zero-temperature one. The LTE at $T=0$ is a
generally valid identity which holds for any individual diagram
constructed with propagators satisfying certain simple properties. For
instance, for the scalar theory one can define the following rules: 

\begin{figure}[h]
\vspace{4mm}
\epsfxsize=6cm
\centerline{\epsffile{cutko.eps}}
\vspace{4mm}
\end{figure}

\begin{figure}[h]
\vspace{3mm}
\epsfxsize=4cm
\centerline{\epsffile{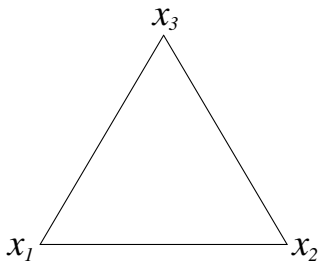}}
\caption{{\it A one loop triangle diagram.}}
\label{fig20}
\vspace{3mm}
\end{figure}

\noindent The `$*$' here means complex conjugation. In addition to
these rules, each {\em 1st type vertex} has a factor $-ig$ and each {\em
2nd type vertex} has a factor $ig$. Using this prescription, we can
construct diagrams in configuration space.  With each diagram then can be
associated a function $F(x_{1},\ldots x_{n})$ having all the 2nd type
vertices underlined.  For example, for the triangle diagram in
Fig.\ref{fig20} we have

\begin{eqnarray}
F(x_{1},x_{2},x_{3}) &=& 
(-ig)^{3} \, i\Delta_{F}(x_{1}-x_{2}) \, i\Delta_{F}(x_{1}-x_{3}) 
\, i\Delta_{F} (x_{2}-x_{3})\nonumber\\
F({\underline{x_{1}}},x_{2},x_{3}) &=&
(-ig)^{2}(ig) \, i\Delta^{+}(x_{1}-x_{2}) 
\, i\Delta^{+}(x_{1}-x_{3})  \, i\Delta_{F} (x_{2}-x_{3})\nonumber\\
F({\underline{x_{1}}},{\underline{x_{2}}},x_{3}) &=&
(ig)^{2}(-ig) 
\, (-i)\Delta_{F}^{*}(x_{1}-x_{2}) \, i\Delta^{+}(x_{1}-x_{3}) \, i\Delta^{+}
(x_{2}-x_{3})\nonumber\\
F({\underline{x_{1}}},{\underline{x_{2}}},{\underline{x_{3}}}) &=&
(ig)^{3} 
\, 
(-i)\Delta_{F}^{*}(x_{1}-x_{2}) 
\, (-i)\Delta_{F}^{*}(x_{1}-x_{3}) \, (-i)\Delta_{F}^{*}
(x_{2}-x_{3}).\nonumber\\
\mbox{etc.}\mbox{\hspace{1.3cm}}&&\nonumber
\end{eqnarray}

\noindent The LTE then states that for a function $F(x_{1}, \ldots, x_{n})$ 
corresponding to some diagram with $n$ vertices

\begin{equation} F( \ldots, {\underline{x_{i}}}, \ldots ) + F( \ldots,
x_{i}, \ldots )= 0, \tseleq{LTE20} \end{equation}

\vspace{3mm}

\noindent provided that $x_{i0}$ is the largest time and all other
underlinings in $F$ are the same. The proof of Eq.(\tseref{LTE20}) is
based on an observation that the propagator $i\Delta_{F}(x)$ can be
decomposed into positive and negative energy parts, i.e. 

\begin{eqnarray}
i\Delta_{F}(x) &=& \theta(x_{0})i\Delta^{+}(x) + 
\theta(-x_{0})i\Delta^{-}(x),\\
i\Delta^{\pm}(x) &=& \int \frac{d^{4}k}{(2\pi)^{3}}e^{-ikx}\theta(\pm 
k_{0})\delta(k^{2}-m^{2}). \tseleq{decomp} \end{eqnarray}

\noindent Incidentally, for $x_{i0}$ being the largest time this 
directly implies

\vspace{-3mm}
\begin{eqnarray}
i\Delta_{F}(x_{j}-x_{i}) &=& 
i\Delta^{-}(x_{j}-x_{i}),\nonumber\\
-i\Delta_{F}^{*}(x_{i}-x_{j}) &=& 
i\Delta^{-}(x_{i}-x_{j}),\nonumber\\
i\Delta_{F}(0) &=& -i\Delta_{F}^{*}(0). 
\tselea{decomp1}
\end{eqnarray}

\noindent As $F(\ldots , {\underline{x_{i}}}, \ldots)$ differs from 
$F(\ldots , x_{i}, \ldots)$ only in the propagators 
directly connected to $x_{i}$ - which are equal (see 
Eq.(\tseref{decomp1})) - and in the sign of the $x_{i}$ vertex, they must 
mutually cancel.  

\vspace{2mm}

\noindent Summing up Eq.(\tseref{LTE20}) for all possible underlinings
(excluding $x_{i}$), we get the LTE where the special r{\^{o}}le of the
largest time is not manifest any more, namely

\begin{equation}
\sum_{index}F(x_{1}, x_{2}, \ldots , x_{n}) = 0.\tseleq{LTE19}\end{equation} 

\vspace{2mm}

\noindent The sum $\sum_{index}$ means summing over all possible
distributions of indices 1 and 2 (or equivalently over all possible
underlinings). The zero-temperature LTE can be easily reformulated for the
$T$-matrices. Let us remind that the Feynman diagrams for the $S$-matrix
($S=1+iT$) can be obtained by multiplying the corresponding $F(x_{1},
\ldots, x_{n})$ with the plane waves for the incoming and outgoing
particles, and subsequently integrate over $x_{1}\ldots x_{n}$. Thus, in
fixed volume quantization a typical $n$-vertex Feynman diagram is given by

\begin{equation}
 \int 
\prod_{i=1}^{n}dx_{i} \prod_{j}
\frac{e^{-ip_{j}x_{m_{j}}}}{\sqrt{2{\omega}_{p_{j}}V}} \prod_{k}
\frac{e^{iq_{k}x_{m_{k}}}}{\sqrt{2{\omega_{q_{k}}}V}}F(x_{1}, \ldots, 
x_{n}). \tseleq{Tmatrix1}
\end{equation}

\vspace{2mm}

\noindent Here the momenta $\{ p_{j}\}$ are attached to incoming particles
at the vertices $\{x_{m_{j}}\}$, while momenta $\{q_{k}\}$ are attached to
outgoing particles at the vertices $\{x_{m_{k}}\}$. In order to
distinguish among various functions $F(x_{1}, \ldots, x_{n})$ with the
same variables $x_{1}, \ldots, x_{n}$, we shall attach a subscript $l_{n}$
to each function $F$. For instance, the function $F_{1_{4}}(x_{1}, \ldots,
x_{4})$ corresponding to the diagram

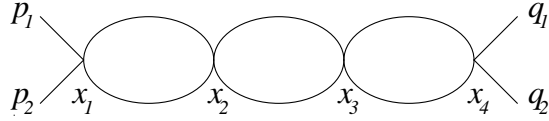
\begin{figure}[h]
\vspace{4mm}
\epsfxsize=7cm
\centerline{\epsffile{fig22.eps}}
\vspace{4mm}
\end{figure}

\vspace{-2mm}

\noindent contributes to $\langle q_{1}q_{2}| iT | p_{1}p_{2}\rangle$ by

\vspace{2mm}

\begin{eqnarray}
\int \prod_{i=1}^{4}dx_{i} \frac{e^{-i(p_{1}+p_{2})x_{1}}}{V\sqrt{4 
\omega_{p_{1}} \omega_{p_{2}}}} \frac{e^{i(q_{1}+q_{2})x_{4}}}{V \sqrt{4 
\omega_{q_{1}} 
\omega_{q_{2}}}}(i\Delta_{F}(x_{1}-x_{2}))^{2}(i\Delta(x_{2}-x_{3}))^{2}
(i\Delta(x_{3}-x_{4}))^{2},\nonumber
\end{eqnarray}

\vspace{2mm}

\noindent similarly, the function $F_{2_{4}}(x_{1}, \ldots, x_{4})$ 
corresponding to the diagram

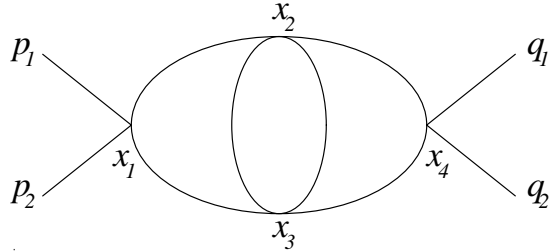
\begin{figure}[h]
\epsfxsize=7cm
\centerline{\epsffile{fig23.eps}}
\end{figure}

\vspace{3mm}

\noindent contributes to $\langle q_{1}q_{2}| iT | p_{1}p_{2}\rangle$ by

\begin{eqnarray}
&&\int \prod_{i=1}^{4}dx_{i} \frac{e^{-i(p_{1}+p_{2})x_{1}}}{V\sqrt{4
\omega_{p_{1}} \omega_{p_{2}}}} \frac{e^{i(q_{1}+q_{2})x_{4}}}{V \sqrt{4
\omega_{q_{1}}
\omega_{q_{2}}}}i\Delta_{F}(x_{1}-x_{2})i\Delta_{F}(x_{1}-x_{3})
(i\Delta_{F}(x_{2}-x_{3}))^{2}\nonumber\\
&&\hspace{5.3cm}\times ~i\Delta_{F}(x_{4}-x_{3})i\Delta_{F}
(x_{4}-x_{2}),\nonumber\\
&&\mbox{ etc.}\nonumber
\end{eqnarray}

\noindent This can be summarized as

\begin{equation}
\langle \{q_{k}\} | iT | \{p_{j}\} \rangle = \sum_{n} \int \ldots \int
\prod_{i=1}^{n}dx_{i} \sum_{l_{n}}\prod_{j}
\frac{e^{-ip_{j}x_{m_{j}}}}{\sqrt{2{\omega}_{p_{j}}V}} \prod_{k}
\frac{e^{iq_{k}m_{n_{k}}}}{\sqrt{2{\omega_{q_{k}}}V}}F_{l_{n}}(x_{1}, 
\ldots, x_{n}), \tseleq{Tmatrix}
\end{equation}

\vspace{2mm}

\noindent Consider now the case $|\{p_{j}\} \rangle = |\{q_{k}\} \rangle$
(let us call it $| a \rangle $). From the unitarity condition: 
$T-T^{\dag} = iT^{\dag}T$, we get

\begin{equation}
\langle a|T |a \rangle - \langle a|T|a \rangle^{*} = i \langle a| 
T^{\dag}T|a \rangle. \tselea{LTE45} \end{equation}

\vspace{2mm}

\noindent On the other hand, by construction $F({\underline{x_{1}}}, \ldots,
{\underline{x_{n}}})= F^{*}(x_{1}, \ldots, x_{n})$, and thus (see
(\tseref{LTE19}))

\begin{equation}
F(x_{1}, \ldots, x_{n}) + F^{*}(x_{1}, \ldots, x_{n}) = - 
\sum_{index^{'}}F(x_{1}, \ldots, x_{n}).
\tseleq{CE1}
\end{equation}

\noindent The prime over {\em index} in (\tseref{CE1}) indicates that we 
sum neither over diagrams with all type 1 vertices nor diagrams with all  
type 2 vertices. Using (\tseref{Tmatrix}), and identifying  
$|\{q_{k}\}\rangle$ with $|\{p_{k}\}\rangle$ ($ = |a \rangle$) we get

\begin{equation}
\langle a|T|a \rangle - \langle a|T|a \rangle^{*} = - \sum_{index^{'}} 
\langle a|T|a \rangle,
\tseleq{CE4}
\end{equation}

\vspace{-3mm}

\noindent or (see (\tseref{LTE45}))
\vspace{-2mm}

\begin{equation}
\langle a |T^{\dag}T| a \rangle = i\sum_{index^{'}}
\langle a |T| a \rangle. \tseleq{CE5}
\end{equation}
\vspace{2mm}

\noindent Eq.(\tseref{CE4}) is the special case of the LTE for the
$T$-matrices. The finite-temperature extension of (\tseref{CE5}) will prove
crucial in Section 4. 

\vspace{2mm}

\begin{figure}[h]
\epsfxsize=9.5cm
\centerline{\epsffile{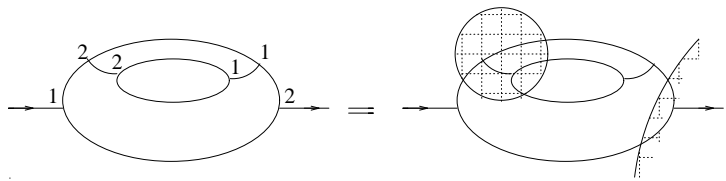}}
\caption{{\it An example of a cut
diagram in the $\varphi ^{3}$ theory which does not contribute to the RHS's
of (\tseref{CE4})-(\tseref{CE5}).  Arrows indicate the flow of
energy.}}
\label{fig.2} 
\end{figure}

\noindent Owing to the $\theta(\pm k_{0})$ in $\Delta^{\pm}(x)$ (see
Eq.(\tseref{decomp})), energy is forced to flow only towards type 2
vertices. From both the energy-momentum conservation in each vertex and
from the energy flow on the external lines, a sizable class of the diagrams
on the RHS's of (\tseref{CE4})-(\tseref{CE5}) will be automatically zero. 
Particularly regions of either 1st or 2nd type vertices which are not
connected to any external line violate the energy conservation and thus do
not contribute (no islands of vertices), see Fig.\ref{fig.2}.
Consequently, the only surviving diagrams are those whose any 1st type
vertex area is connected to incoming particles and any 2nd type vertex
area is connected to outgoing ones. From historical reasons the border
between two regions with different type of vertices is called {\em cut}
and corresponding diagrams are called {\em cut diagrams}. 

\vspace{3mm}

\begin{figure}[h]
\vspace{3mm}
\epsfxsize=5cm 
\centerline{\epsffile{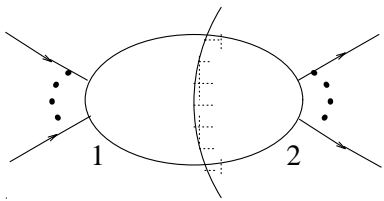}}
\caption{{\it Generic form of the cut diagram at the $T=0$. Shadow is on the
2nd type vertex area. }} \label{fig4}
\vspace{3mm}
\end{figure}

\noindent We have just proved a typical feature of $T=0 $ QFT, namely
any cut diagram is divided by the cut into two areas only, see
Fig.\ref{fig4}. Eq.(\tseref{CE4}), rewritten in terms of the cutings is so
called {\em cutting equation} (or Cutkosky's cutting rules) \tsecite{PN,
V, TV}.

\vspace{2mm}

\noindent One point should be added. Inserting the completeness relation
$\sum_{f} |f\rangle \langle f| =1$ into the LHS of (\tseref{CE5}), we get

\begin{equation}
\sum_{f}\langle a|T^{\dag}|f \rangle \langle f|T|a \rangle = i 
\sum_{cuts} \langle a|T|a \rangle.
\tselea{MLTE}
\end{equation}

\noindent It may be shown \tsecite{PL2, PN} that all the intermediate
particles in $|f \rangle$ correspond to cut lines. This has a natural
extension when $\langle a|T^{\dag}T|a \rangle \rightarrow \langle
a|T^{\dag}{\cali{P}}T|a \rangle$ with ${\cali{P}}$ being a projection
operator (${\cali{P}}={\cali{P}}^{\dag}={\cali{P}}^{2}$) which eliminates
some of the states $|f \rangle$. It is easy to see that in such case

\begin{equation}
\langle a|T^{\dag}{\cali{P}}T| a \rangle = i{\tilde{\sum_{cuts}}} 
\langle a|T|a \rangle, 
\tselea{MLTE1}
\end{equation}

\noindent where tilde over the $\sum_{cuts}$ indicates that one sums over 
the diagrams which do not have the cut lines corresponding to particles 
removed by ${\cali{P}}$. 
  
\vspace{2mm}

\noindent There is no difficulty in applying the previous results to
spin-$\frac{1}{2}$ \tsecite{PN, V}. The LTE follows as before: the diagram
with only $iS_{F}$ propagators (and $-ig$ per each vertex) plus the
diagram with only $\hat{(iS_{F})}$ propagator {\footnote{The function
$i{\hat{S}}_{F}(x)$, similarly as $(i\Delta_{F})^{*}(x)$, interchanges the
r{\^{o}}le $S^{+}$ and $S^{-}$. Unlike bosons, for fermions
$i{\hat{S}}_{F}(x)$ is not equal to $(iS_{F})^{*}(x)$. Despite that, 
Eq.(\tseref{CE4}) still holds \tsecite{PN}.}} (and $ig$ per
each vertex) equals to minus the sum of all diagrams with one up to $n-1$
the type 2 vertices ($n$ being the total number of vertices). For gauge
fields more care is needed. Using the Ward identities one can show
\tsecite{PN} that type 1 and type 2 vertices in
(\tseref{CE4})-(\tseref{CE5}) may be mutually connected only by {\em
physical particle} propagators (i.e. neither through the propagators
corresponding to particles with non-physical polarizations or Fadeev-Popov
ghosts and antighosts).

\subsection{Thermal Wick's theorem (the Dyson-Schwinger equation)}

The key observation at finite temperature is that for systems of {\em
non-interacting} particles in thermodynamical equilibrium Wick's theorem is
still valid, i.e. one can decompose the $2n$-point (free) thermal Green
function into a product of two-point (free) thermal Green functions. This
may be defined recursively by

\begin{equation} \langle {\cali{T}} ( \psi (x_{1}) \ldots \psi (x_{2n}) )
{\rangle}_{\beta}= \sum_{\stackrel{j}{j \not= i}} \varepsilon_{P}\langle
{\cali{T}} ( \psi (x_{i})\psi (x_{j}) ) {\rangle}_{\beta} ~\langle
{\cali{T}} ( \prod_{k \not= i;j}\psi(x_{k})) {\rangle}_{\beta},
\tseleq{wick} \end{equation} 

\noindent where $\varepsilon_{P}$ is the signature of the permutation of
fermion operators ($=1$ for boson operators) and ${\cali{T}}$ is the 
standard time ordering symbol. We shall use, from now on,
the subscript $\beta$, emphasizing that the thermal mean value describes a
system in thermodynamical equilibrium (at the temperature $\beta^{-1}$).
Note that the choice of ``$i$" in (\tseref{wick}) is completely arbitrary. 
The proof can be found for example in \tsecite{LB,Mills,Evans}.  Similarly
as at $T=0$, Wick's theorem can also be written for the (free) thermal
Wightman functions \tsecite{Mills, IZ}, i.e. 

\begin{equation}
\langle \psi (x_{1}) \ldots \psi (x_{2n})
{\rangle}_{\beta}=
\sum_{\stackrel{j}{j \not= 1}} \varepsilon_{P}\langle \psi (x_{1})\psi 
(x_{j}) {\rangle}_{\beta} ~  \langle
\prod_{k \not= 1;j}\psi(x_{k}) \rangle_{\beta}.
\tseleq{wick2}
\end{equation}

\noindent A particularly advantageous form of this is the so called
Dyson-Schwinger equation (see Appendix A) which, at the $T \not= 0$, reads

\begin{equation} \langle G[\psi] \psi(x) F[\psi] \rangle_{\beta} = \int dz
\langle \psi(x) \psi(z) \rangle_{\beta} \left\langle G[\psi] 
\frac{{\stackrel{\rightarrow}{\delta}} 
F[\psi] }{\delta \psi(z)} \right\rangle_{\beta} + \int dz \langle \psi(z)
\psi(x) \rangle_{\beta} \left\langle \frac{ 
G[\psi]\stackrel{\leftarrow}{\delta}}{\delta 
\psi(z)} F[\psi] \right\rangle_{\beta}, \tseleq{S-D2} \end{equation}

\vspace{2mm}

\noindent where $\psi(x)$ is an interaction-picture field and $G[\ldots]$ and
$F[\ldots]$ are functionals of $\psi$. The arrowed
variations $\frac{\delta}{\delta \psi(z)}$ are defined as a formal
operation satisfying two conditions, namely:

\begin{equation}
\frac{\stackrel{\rightarrow}{\delta}}{\delta \psi_{n}(z)}(\psi_{m}(x) 
\psi_{q}(y)) = 
\frac{\delta \psi_{m}(x)}{\delta 
\psi_{n}(z)}\psi_{q}(y)+(-1)^{p}\psi_{m}(x)\frac{\delta 
\psi_{q}(y)}{\psi_{n}(z)}, \nonumber
\tseleq{var1}
\end{equation}

\noindent or

\vspace{-2mm}

\begin{equation} 
(\psi_{m}(x) \psi_{q}(y))\frac{\stackrel{\leftarrow}{\delta}}{\delta 
\psi_{n}(z)} = (-1)^{p}\frac{\delta \psi_{m}(x)}{\delta 
\psi_{n}(z)}\psi_{q}(y)+\psi_{m}(x)\frac{\delta \psi_{q}(y)}{\psi_{n}(z)},
\nonumber
\tseleq{var12}
\end{equation}

\noindent with

\vspace{-2mm}

\begin{equation}
\frac{\delta \psi_{m}(x)}{\delta \psi_{n}(y)} = 
\delta(x-y)\delta_{mn}.\nonumber \tseleq{var2}
\end{equation}

\vspace{2mm} 

\noindent The ``$p$'' is $0$ for bosons and $1$ for fermions; subscripts
$m,n$ suggest that several types of fields can be generally present. Note,
for bosons $\frac{{\stackrel{\rightarrow}{\delta}}F}{\delta \psi}
=\frac{F\stackrel{\leftarrow}{\delta}}{\delta \psi}$ which we shall denote
as $\frac{\delta F}{\delta \psi}$. For more details see Appendix A.

\subsection{Thermal largest-time equation}

The LTE (\tseref{CE5}) can be extended to the finite-temperature case,
too.  Summing up in (\tseref{CE5}) over all the eigenstates of $K$ ($= H-\mu
N$) with the weight factor $e^{-\beta K_{i}}$ ($i$ labels the
eigenstates), we get

\begin{equation}
\langle TT^{\dag} \rangle_{\beta} = i \sum_{index^{'}} \langle T
\rangle_{\beta}.
\tseleq{TLTE2}
\end{equation}

\vspace{2mm}

\noindent Let us consider the RHS of (\tseref{TLTE2}) first. The
corresponding thermal LTE and diagrammatic rules (Kobes-Semenoff rules
\tsecite{LB}) can be derived precisely the same way as at $T=0$ using
the previous, largest-time argumentation \tsecite{LB, KS}. It turns out
that these rules have basically identical form as those in the previous
Section, with an exception that now $\langle 0 | \ldots |0 \rangle
\rightarrow \langle \ldots \rangle_{\beta}$. Note that labeling vertices
by 1 and 2 we have naturally got a doubling of the number of degrees of
freedom. This is a typical feature of the {\em real-time formalism} in
thermal QFT (here, in so called {\em Keldysh version} \tsecite{LB}). 

\vspace{2mm}

\noindent We should also emphasize that it may happen some fields are not
thermalized. For example, external particles entering a heat bath or
particles describing non-physical degrees of freedom \tsecite{LR}.
Particularly, if some particles (with momenta $\{ p_{j} \}$) enter the
heat bath, the mean statistical value of an observable $A$ is then

\begin{eqnarray}
\sum_{i} \frac{e^{-\beta K_{i}}}{Z} \langle i; \{ p_{j} \}|A|i; \{ p_{j} 
\} \rangle &=& Z^{-1} Tr(\rho_{\{ p_{j} \}} \otimes e^{-\beta K} 
A),\nonumber\\
\rho_{\{ p_{j} \}} &=& |\{ p_{j} \}\rangle \langle \{ p_{j} \} |,\nonumber
\end{eqnarray}

\noindent which we shall denote as $\langle A \rangle_{\{ p_{j} \}, \beta}$. 
From this easily follows the generalization of (\tseref{TLTE2}) 

\begin{equation}
\langle TT^{\dag} \rangle_{\{p_{k}\}, \beta} = i \sum_{index^{'}} \langle T
\rangle_{\{p_{k}\}, \beta}. \tseleq{TLTE3}
\vspace{1mm}
\end{equation}

\noindent Unlike $T=0$, we find that the cut diagrams have disconnected
vertex areas and no kinematic reasonings used in last Section can, in
general, get rid of them. This is because the thermal part of $\langle
\varphi(x) \varphi(y) \rangle_{\beta}$ \footnote{Note that $\langle
\varphi(x) \varphi(y) \rangle _{\beta} = \langle : \varphi(x) \varphi(y):
\rangle_{\beta} + \langle 0| \varphi(x) \varphi(y) |0 \rangle $ and
$\langle : \varphi(x) \varphi(y) : \rangle_{\beta} = \int
\frac{d^{4}k}{(2\pi)^{3}}f_{B}(k_{0}) \delta (k^{2}-m^{2}) e^{-ik(x-y)}$, 
with $f_{B}(k_{0})=(e^{\beta |k_{0}|} -1)^{-1}. $}
describes the absorption of on shell particle from the heat bath or the
emission of one into it. Thus, at $T \not = 0 $, there is no definite
direction of transfer of energy from type 1 vertex to type 2 one as energy
flows in both directions. Some cut diagrams nevertheless vanish. It is
simple to see that only those diagrams survive in which the
non-thermalized external particles ``enter'' a diagram via the 1st type
vertices and ``leave'' it via the 2nd type ones. We might deduce this from
the definition of $\langle T \rangle_{\{ p_{j} \}, \beta}$, indeed

\begin{equation}
\sum_{index^{'}} \langle T \rangle_{\{ p_{j} \}, \beta}= \sum_{index^{'}} 
\sum_{i} \frac{e^{-\beta K_{i}}}{Z}
\langle i;\{ p_{j} \}|T|i; \{ p_{j} \} \rangle.\nonumber
\end{equation}

\noindent Note, we get the same set of thermal cut diagrams interchanging
the summation $\sum_{index^{'}}$ with $\sum_{i}$. It is useful to start
then with $\sum_{index^{'}} \langle i;\{ p_{j} \}|T|i; \{ p_{j} \}
\rangle$.  This is, as usual, described by the ($T=0$) cutting rules. In
the last Section we learned that the general structure of the corresponding
cut diagrams is depicted in Fig.\ref{fig4}, particularly the external
particles enter the cut diagram via type 1 vertices and leave it via type
2 ones. Multiplying each diagram (with the external particles in the state
$|i; \{p_{j}\} \rangle$) with the prefactor $\frac{e^{-\beta K_{i}}}{Z}$
and summing subsequently over $i$, we again retrieve the thermal cut
diagrams, though now it becomes evident that the particles $\{ p_{j} \}$
enter such diagram only via type 1 vertices and move off only through type
2 ones, since the summation of the ($T=0$) cut diagrams from which it was
derived does not touch lines corresponding to unheated particles. Note,
the latter analysis naturally explains why the unheated particles obey the
($T=0$) LTE diagrammatic rules even in the thermal diagrams

\begin{figure}[h]
\epsfxsize=15.3cm
\centerline{\epsffile{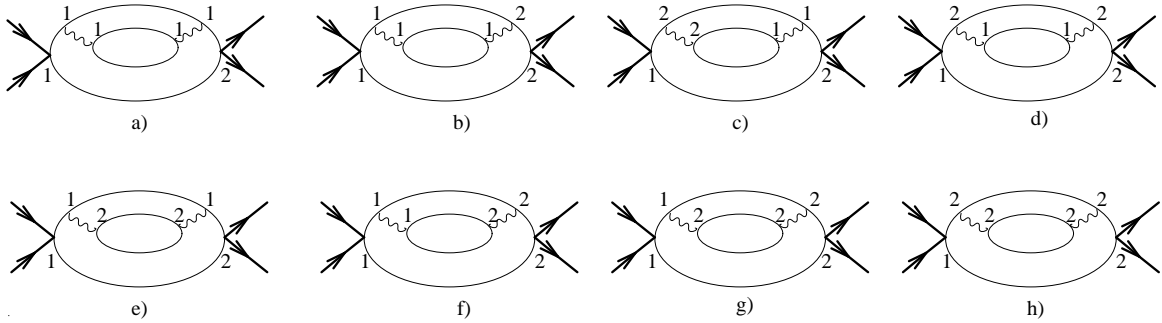}}
\caption{{\it An example of non-vanishing cut diagrams at the $T\not =0$.
The heat-bath consists of two different particles. External particles are
not thermalized.}}
\label{fig.3}
\end{figure}

\vspace{2mm} \noindent Another vanishing comes from kinematic reasons.
Namely three-leg vertices with all the on shell particles (1-2 lines) can
not conserve energy-momentum and consequently the whole cut diagram is
zero.  As an illustration let us consider all the non-vanishing,
topologically equivalent cut diagrams of given type involved in a
three-loop contribution to $i\sum_{index^{'}} \langle T
\rangle_{pq,\beta}$ (see Fig.\ref{fig.3}).  \footnote{Let us emphasize
that originally we had the 64 possible cut diagrams.}

\vspace{2mm}

\begin{figure}[h] \vspace{3mm} \epsfxsize=15.3cm
\centerline{\epsffile{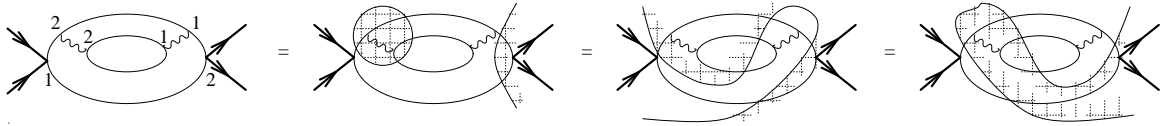}} \caption{{\it The cut diagram from
Fig.\ref{fig4} c) demonstrates that the cut can be defined in many ways but
the number of crossed lines is still the same.}} \label{fig6} \vspace{3mm}
\end{figure}

\noindent Let us stress one more point. In contrast with $T=0$, at finite
temperature the cut itself neither is unique nor defines topologically
equivalent areas, see Fig.\ref{fig6}, only the number of crossed legs is,
by definition, invariant. 

\vspace{2mm}

\noindent This ambiguity shows that the concept of the cut is not very useful
at finite temperature and in the following we shall refrain from using it. 

\vspace{3mm}

\noindent In Chapter 4 it will prove useful to have an analogy of
(\tseref{TLTE3}) for $\langle T^{\dag}{\cali{P}}T \rangle_{\beta}$. Here
$\cali{P}$ is the projection operator defined as

\begin{equation}
{\cali{P}} = \sum_{j} |a;j\rangle  \langle a;j|,
\tseleq{proj}
\vspace{1mm}
\end{equation}

\noindent where ``$j$'' denotes the physical states for the heat-bath
particles and ``$a$'' labels the physical states for the outgoing, 
non-thermalized particles. Let us deal with $\langle T^{\dag}{\cali{P}}T 
\rangle_{\beta}$. Using (\tseref{MLTE1}), we acquire

\begin{equation} \langle T^{\dag}{\cali{P}}T\rangle_{\beta} = i\sum_{l}
\frac{e^{-\beta K_{l}}}{Z} {\tilde{\sum_{index^{'}}}} \langle l|T|l
\rangle. \end{equation}

\noindent Interchanging the summations, we finally arrive at

\begin{equation}
\langle T^{\dag} {\cali{P}} T \rangle_{\beta} = i{\tilde{\sum_{index^{'}}}}
\langle T \rangle_{\beta},
\tseleq{proj2}
\vspace{1mm}
\end{equation}

\noindent where tilde over the $\sum_{index^{'}}$ means that we are
restricted to consider the cut diagrams, with only (1-2)-particle lines
corresponding to the ``$a$'' and ``$j$''particles (i.e $ \langle 0|
\varphi(x) \varphi(y) |0 \rangle$ and $\langle \psi(x) \psi(y)
\rangle_{\beta}$, respectively). The extension of Eq.(\tseref{proj2}) to
the case where some external, non-thermalized particles $\{p_{k}\}$ are
present is obvious, and reads

\begin{equation}
\langle T^{\dag} {\cali{P}} T \rangle_{\{p_{k}\}, \beta} =
i{\tilde{\sum_{index^{'}}}} \langle T \rangle_{\{p_{k}\},
\beta}.
\tseleq{proj4}
\end{equation}

\vspace{1mm}

\noindent Finally, let us note that using the LTE, one may extend the
previous treatment to various Green functions. The LTE
for Green's functions is then a useful starting point for dispersion
relations, see e.g.  \tsecite{LB, KS}.

\section{Heat-Bath particle number spectrum:\\
 general framework}

The cutting equation (\tseref{proj4}) can be fruitfully used for both the
partition function $Z$ and the heat-bath particle number spectrum
$\frac{d\langle N(\omega) \rangle}{d \omega} $ calculations. To see that,
let us for simplicity assume that two particles (say $\Phi_{1}, \Phi_{2}$)
scatter inside a heat bath. We are interested in the heat-bath number
spectrum after two different particles (say $\phi_{1}, \phi_{2})$ appear in
the final state. Except for the condition that the external particles are
different from the heat bath ones, no additional assumption about their
nature is needed at this stage. 

\vspace{2mm}

\noindent The initial density matrix $\rho_{i}$ (i.e. the density matrix
describing the physical situation before we introduce the particles
$\Phi_{1}(p_{1}), \Phi_{2}(p_{2})$ into the oven) can be written as

\begin{equation} \vspace{2mm} \rho_{i} = Z_{i}^{-1} \sum_{j} e^{-\beta
K_{j}}|j;p_{1}, p_{2} \rangle \langle j; p_{1}, p_{2}|, \tseleq{rhoi}
\end{equation}

\noindent where ``$j$'' denotes the set of occupation numbers for the heat
bath particles. A long time after the scattering the final
density matrix $\rho_{f}$ reads 

\begin{equation}
\rho_{f} = Z_{f}^{-1} \sum_{j} e^{-\beta K_{j}} {\cali{P}}S|j; p_{1}, p_{2}
\rangle \langle j; p_{1}, p_{2}|S^{\dag} {\cali{P}}^{\dag},
\tseleq{rhof}
\vspace{1mm}
\end{equation}

\noindent here ${\cali{P}}$ is the projection operator projecting out all
the non-heat-bath final states except of $\phi_{1}(q_{1}),
\phi_{2}(q_{2})$ ones. The $S$-matrix in (\tseref{rhof}) is defined in a
standard way: $S=1+iT$. The $Z_{f}$ in (\tseref{rhof}) must be different
from $Z_{i}$ as otherwise $\rho_{f}$ would not be normalized to unity. In
order that $\rho_{f}$ satisfy the normalization condition $Tr(\rho_{f}) =
1$, one finds

\begin{equation}
\vspace{1mm}
Z_{f}= \sum_{j} e^{-\beta K_{j}} \langle j; p_{1}, p_{2}|S^{\dag} {\cali{P}}
S|j; p_{1},
p_{2} \rangle = {}\langle S^{\dag}{\cali{P}} S \rangle_{p_{1}
p_{2}, \beta} ~ Z_{i}  = \langle T^{\dag}{\cali{P}} T \rangle_{p_{1}
p_{2}, \beta} ~ Z_{i}. 
\tseleq{Zf} 
\end{equation}

\noindent The key point is that we have used in (\tseref{Zf}) the $T$ -
matrix because the initial state $|\Phi_{1} (p_{1}), \Phi_{2} (p_{2})
\rangle$ is, by definition, different from the final one $|\phi
(q_{1}),\phi_{2} (q_{2}) \rangle$ and consequently ${\cali{P}}S$ can be
replaced by $i{\cali{P}}T$. This allows us to calculate $Z_{f}$ using
directly the diagrammatic technique outlined in the preceding Section. 

\vspace{2mm}

\noindent From (\tseref{mean value}) and (\tseref{rhof}) one can directly 
read off that the number spectrum of the heat bath particles is:

\begin{eqnarray} \frac{d \langle N_{l}( \omega )\rangle_{f}}{ d\omega} & = & 
\int \frac{ d^{3}{\vect{k}}}{(2 \pi)^{3}} {\delta}^{+}(
{\omega}^{2} - {\vect{k}}^{2} - m_{l}^{2})
\sum_{f} \langle f | a^{\dag}_{l}({\vect{k}}; \omega)a_{l}({\vect{k}};
\omega) \rho_{f}| f \rangle \nonumber  \\
 & = & \int \frac{d^{3}{\vect{k}}}{(2 \pi)^{3}} \delta^{+}(
\omega^{2}-{\vect{k}}^{2} -m_{l}^{2}) \frac{{}\langle 
T^{\dag}{\cali{P}}
a^{\dag}_{l}({\vect{k}}; \omega)a_{l}({\vect{k}}; \omega) T
\rangle_{p_{1} p_{2},
\beta}}{{} \langle T^{\dag} {\cali{P}}T \rangle_{p_{1} p_{2}, \beta}},
\tselea{dN}
\end{eqnarray}

\noindent and consequently

\begin{equation} \langle N_{l} \rangle_{f} = \int \frac{d^{4}k}{(2
\pi)^{3}} \delta^{+}(k^{2}-m_{l}^{2})  \frac{{}\langle 
T^{\dag}{\cali{P}}
a^{\dag}_{l}(k)a_{l}(k) T \rangle_{p_{1} p_{2},
\beta}}{{} \langle T^{\dag} {\cali{P}}T \rangle_{p_{1} p_{2}, \beta}},   
\vspace{1mm}
\tseleq{N}
\end{equation}

\noindent where we have used the completeness relation for the final
states $|f \rangle$ and $[{\cali{P}};a^{\dag}a]=0$. The subscript ``$l$''
denotes which type of heat-bath particles we are interested in. In the
following the index will be mostly suppressed.

\section{Modified cut diagrams}

To proceed further with (\tseref{dN}) and (\tseref{N}), we expand the 
$T$-matrix in terms of time-ordered  interaction-picture fields, i.e.

\begin{equation} T[\psi] = \sum_{n} \int dx_{1} \ldots \int dx_{n}
\alpha_{n}(x_{1}\ldots x_{n}) {\cali{T}}(\psi(x_{1})\ldots \psi(x_{n})). 
\tseleq{Hin} \vspace{1mm} \end{equation}

\noindent Here $\psi$ represents a heat-bath field in the interaction
picture. Other fields (i.e. ${\overline{\phi}}$,$\phi$ and $\Phi$) are
included in the $\alpha_{n}$ \footnote{ When Fermi fields are involved, we
have, for the sake of compactness, included in the argument of $\psi$ the
space-time coordinate, the Dirac index, and a discrete index which
distinguishes $\psi_{\alpha}$ from ${\overline{\psi}}_{\alpha}$.}. An
extension of (\tseref{Hin}) to the case where different heat-bath fields
are present is natural. Employing (\tseref{Hin}) in $\langle
T^{\dag}{\cali{P}} T\rangle_{p_{1}p_{2},\beta}$, one can readily see that
this factorizes out in each term of the expansion a {\em pure} thermal
mean value $\langle \ldots \rangle_{\beta}$. The general structure of each
such thermal mean value is: ${\langle G_{m}[\psi] F_{n}[\psi]
\rangle}_{\beta}$, where $F_{n}[\ldots]$ and $G_{m}[\ldots]$ are the
operators with ``$n$" chronological and ``$m$" anti-chronological time
ordered (heat-bath) fields, respectively. Analogous factorization is true
in the expansion of $~{} \langle T^{\dag}{\cali{P}}^{\dag}a^{\dag} a
{\cali{P}} T \rangle_{p_{1}p_{2}, \beta}$. The only difference is that the
pure thermal mean value has the form $\langle G_{m}[\psi] a^{\dag}a
F_{n}[\psi] \rangle_{\beta}$ instead \footnote{Remember that ${\cali{P}}=
{\cali{P}}^{'} \otimes {\cali{P}}^{"} = |q_{1}, q_{2}\rangle \langle
q_{1}, q_{2}| \otimes \sum_{j}|j\rangle \langle j|$. Here ${\cali{P}}^{"}=
\sum_{j} |j \rangle \langle j|$ behaves as an identity in the subspace of
heat-bath states.}. In case when various heat-bath fields are present, $m
= m_{1}+m_{2}+ \ldots +m_{n}$, with ``$m_{l}$'' denoting the number of the
heat-bath fields of $l$-th type. 

\vspace{2mm}

\noindent Applying the Dyson-Schwinger equation to $\langle
G_{m}[\psi]a^{\dag}a F_{n}[\psi] \rangle_{\beta}$ twice and summing over 
``$n$" and ``$m$", we get cheaply the following expression (c.f. also  
(\tseref{S-D8}))

\vspace{-2mm}
\begin{eqnarray}
\lefteqn{\langle T^{\dag}{\cali{P}} a_{l}^{\dag}a_{l} T
\rangle_{p_{1}p_{2},\beta} =}\nonumber \\ &= & \int
dxdy \{ \langle \psi_{l}(x)a_{l}^{\dag} \rangle_{\beta} \langle
a_{l} \psi_{l}(y) \rangle_{\beta} + (-1)^{p}\langle \psi_{l}(x) a_{l}
\rangle_{\beta} \langle a_{l}^{\dag} \psi_{l}(y) \rangle_{\beta}
\} \left\langle
\frac{ 
T^{\dag}{\stackrel{\leftarrow}{\delta}}}{\delta 
\psi_{l}(x)}{\cali{P}} \frac{{\stackrel{\rightarrow}{\delta}} 
T}{\delta \psi_{l}(y)} \right\rangle_{p_{1}p_{2}, \beta} \nonumber \\
&+& \int
\frac{dxdy}{2}\{\langle \psi_{l}(x)a_{l} \rangle_{\beta}\langle
\psi_{l}(y)a_{l}^{\dag} \rangle_{\beta} +(-1)^{p} \langle \psi_{l}(x) a_{l}^{\dag}
\rangle_{\beta}\langle \psi_{l}(y)a_{l}  
\rangle_{\beta}\} \left\langle
\frac{T^{\dag}{\stackrel{\leftarrow}{\delta^{2}}}}{\delta \psi_{l}(y) \delta 
\psi_{l}(x)}{\cali{P}}T
\right\rangle_{p_{1} p_{2}, \beta} \nonumber\\ &+& \int
\frac{dxdy}{2}\{\langle a_{l}\psi_{l}(x) \rangle_{\beta} \langle
a_{l}^{\dag} \psi_{l}(y) \rangle_{\beta} + (-1)^{p} \langle a_{l}^{\dag} 
\psi_{l}(x) \rangle_{\beta} \langle a_{l}\psi_{l}(y)
\rangle_{\beta} \} \left\langle
T^{\dag} {\cali{P}}\frac{{\stackrel{\rightarrow}{\delta^{2}}}T}{\delta 
\psi_{l}(y) \delta \psi_{l}(x)} \right\rangle_{p_{1} p_{2}, \beta} 
\nonumber\\
&+& \langle a_{l}^{\dag}a_{l} \rangle_{\beta} \langle T^{\dag}{\cali{P}}T
\rangle_{p_{1}p_{2}, \beta}, \tselea{average1}
\end{eqnarray}

\noindent A similar decomposition for $\langle T^{\dag}{\cali{P}}T
\rangle_{p_{1}p_{2}, \beta}$ would not be very useful
(cf.(\tseref{S-D13})); instead we define $\langle (T^{\dag}{\cali{P}}T)^{'}
\rangle_{p_{1}p_{2}, \beta}$ having the same expansion as $\langle
T^{\dag}{\cali{P}}T \rangle_{p_{1}p_{2}, \beta}$ except for the
$\alpha_{n}(\ldots){\cali{P}}\alpha_{m}^{\dag}(\ldots)$ are replaced by
$\alpha_{n}(\ldots){\cali{P}}\alpha_{m}^{\dag}(\ldots)\frac{n_{l}+m_{l}}{2}$.
In this formalism $\langle ( T^{\dag}{\cali{P}}T)^{'}\rangle_{p_{1}p_{2},
\beta}$ is

\vspace{-2mm}
\begin{eqnarray}
\lefteqn{\langle (T^{\dag}{\cali{P}} T)^{'} \rangle_{p_{1}p_{2}, 
\beta}}\nonumber\\ &=& \int dxdy {\langle 
\psi_{l}(x)\psi_{l}(y) \rangle}_{\beta} \left\langle
\frac{T^{\dag} {\stackrel{\leftarrow}{\delta}}}{\delta \psi_{l}(x)}{\cali{P}} 
\frac{{\stackrel{\rightarrow}{\delta}} T}{\delta \psi_{l}(y)} 
\right\rangle_{p_{1} p_{2}, \beta} \nonumber\\ &+& \int
\frac{dxdy}{2}{\langle {\overline{\cali{T}}} ( \psi_{l}(x)\psi_{l}(y) )
\rangle}_{\beta} \left\langle
\frac{T^{\dag}{\stackrel{\leftarrow}{\delta^{2}}}}{\delta \psi_{l}(y) \delta 
\psi_{l}(x))}{\cali{P}}T \right\rangle_{p_{1}p_{2}, \beta}
\nonumber\\ &+& \int
\frac{dxdy}{2}\langle {\cali{T}} ( \psi_{l}(x)\psi_{l}(y) )
\rangle_{\beta} \left\langle
T^{\dag} {\cali{P}}\frac{{\stackrel{\rightarrow}{\delta^{2}}}T}{\delta 
\psi_{l}(y) \delta \psi_{l}(x)} \right\rangle_{p_{1}p_{2},\beta},
\tselea{average2}
\end{eqnarray}

\vspace{1mm}

\noindent with the ${\overline{\cali{T}}}$ being the anti-chronological
ordering symbol. Comparing (\tseref{average2}) with (\tseref{S-D12}), we
can interpret the RHS of (\tseref{average2}) as a sum over {\em all}
possible distributions of one line (corresponding to $\psi_{l}$) inside of
each given ($T\not= 0$ !) cut diagram constructed out of $\langle
T^{\dag}{\cali{P}}T \rangle_{p_{1}p_{2}, \beta}$. As (\tseref{average2})
has precisely the same diagrammatical structure as $\langle
T^{\dag}{\cali{P}}a^{\dag}a T \rangle_{p_{1}p_{2}, \beta} -\langle
a^{\dag}a \rangle_{\beta} \langle T^{\dag}{\cali{P}}T \rangle_{p_{1}p_{2},
\beta}$ (cf.(\tseref{average1})), it shows that in order to compute the
numerator of $\frac{d\Delta \langle N(\omega) \rangle}{d\omega} =
\frac{d\langle N(\omega) \rangle_{f}}{d\omega}-\frac{d\langle N(\omega)
\rangle_{i}}{d\omega}$ {\footnote{Here $\frac{d\langle N(\omega)
\rangle_{i}}{d\omega} = \int\frac{d^{3}{\vect{k}}}{(2\pi)^{3}}
\delta^{+}(\omega^{2}-{\vect{k}}^{2} - m^{2}) \langle a^{\dag}(\omega,
{\vect{k}})a(\omega, {\vect{k}}) \rangle_{\beta}$, (cf. (\tseref{dN})).}}
one can simply modify the usual $\langle
T^{\dag}{\cali{P}}T\rangle_{p_{1}p_{2}, \beta}$ cut diagrams by the
following one-line replacements (cf.(\tseref{dN})). 

\vspace{4mm}

\noindent {\bf (i) For neutral scalar bosons:} 

\begin{eqnarray}
\langle  \varphi(x)\varphi(y) \rangle_{\beta} &\rightarrow &
\int \frac{d^{3}{\vect{k}}}{(2\pi)^{3}}\; \delta^{+}(\omega^{2} 
-{\vect{k}}^{2}-m^{2}) \{ 
\langle \varphi(x)a^{\dag}({\vect{k}};\omega) \rangle_{\beta} \langle
a({\vect{k}};\omega ) \varphi(y) \rangle_{\beta}\nonumber\\
 && +\; \langle \varphi(x) 
a({\vect{k}};\omega) \rangle_{\beta} \langle 
a^{\dag}({\vect{k}};\omega) \varphi(y) \rangle_{\beta} \}\nonumber \\
&&\mbox{\vspace{5mm}}\nonumber\\ 
&=& \int
\frac{d^{4}k}{(2\pi)^{3}}\; \delta(k^{2}-m^{2}) \left\{ 
f_{B}(\omega)(f_{B}(\omega)
+ 1) \right. \nonumber \\
&& \times 
~(\delta^{-}(k_{0}+\omega)+\delta^{+}(k_{0}-\omega))\nonumber\\ 
&& +\left. \delta^{+}(k_{0}-\omega)(1 +
f_{B}(\omega)) - \delta^{-}(k_{0}+\omega)f_{B}(\omega) \right\} 
e^{-ik(x-y)},
\tselea{new1}
\end{eqnarray}
\vspace{-1mm}

\noindent where $f_{B}(\omega)$ is the Bose-Einstein distribution: 
$f_{B}(\omega)= \frac{1}{e^{\beta|\omega|}-1}$. Term
$\theta(-k_{0})f_{B}(\omega)$ describes the absorption of a heat-bath
particle, so reduces the number spectrum, that is why the negative sign
appears in front of it. Analogously,

\vspace{-1mm}

\begin{eqnarray}
\langle {\cali{T}} ( \varphi(x)\varphi(y) ) \rangle_{\beta}
&\rightarrow & 
\int \frac{d^{3}{\vect{k}}}{(2\pi)^{3}}\;\delta^{+}(\omega^{2}
-{\vect{k}}^{2}-m^{2}) \{\langle a^{\dag}({\vect{k}};\omega) \varphi(x)
\rangle_{\beta} \langle a({\vect{k}};\omega)\varphi(y)
\rangle_{\beta}\nonumber\\
&&+ \;\langle a({\vect{k}};\omega)\varphi(x) \rangle_{\beta} \langle
a^{\dag}({\vect{k}};\omega) \varphi(y) \rangle_{\beta}\}\nonumber\\
&&\mbox{\vspace{5mm}}\nonumber\\
&=& 
\int \frac{d^{4}k}{(2\pi)^{3}}
\delta(k^{2}-m^{2})(1 +   
f_{B}(\omega))f_{B}(\omega)e^{-ik(x-y)}\nonumber\\
&& \times ~(\delta^{+}(k_{0}-\omega)+\delta^{-}(k_{0}+\omega)).
\tselea{new4}
\end{eqnarray}

\vspace{2mm}

\noindent Similarly, for $\Delta \langle N \rangle$ one needs the
following replacements (cf.(\tseref{N}))

\begin{eqnarray}
\langle  \varphi(x)\varphi(y) \rangle_{\beta} &\rightarrow & \int
\frac{d^{4}k}{(2\pi)^{3}}\;\delta(k^{2}-m^{2})\left\{ 
f_{B}(\omega_{k})(f_{B}(\omega_{k}) + 
1) \right. 
\nonumber \\ &&\left. +~ \theta(k_{0})(1 + f_{B}(\omega_{k})) -
\theta(-k_{0})f_{B}(\omega_{k})\right\}e^{-ik(x-y)}\nonumber,\\
&&\mbox{\vspace{1cm}}\nonumber\\
\langle {\cali{T}} ( \varphi(x)\varphi(y) ) \rangle_{\beta}
&\rightarrow & 
\int \frac{d^{4}k}{(2\pi)^{3}}\;
\delta(k^{2} - m^{2})(1 + 
f_{B}(\omega_{k}))f_{B}(\omega_{k})e^{-ik(x-y)},\nonumber\\
\tselea{new2}
\end{eqnarray}

\vspace{2mm}

\noindent with $\omega_{k}=\sqrt{{\vect{k}}^{2}-m^{2}}$. 

\vspace{1cm}

\noindent {\bf (ii) For Dirac fermions:}

\vspace{3mm} 

\noindent The Dirac field is comprised of two different types of  
excitations (mutually connected via charge conjugation),
so the corresponding number operator $N(\omega) = N_{b}(\omega) +
N_{d}(\omega)$ with

\begin{eqnarray}
N_{b}(\omega) &=& \sum_{\alpha =1,2} \int   
\frac{d^{3}{\vect{k}}}{(2\pi)^{3}}\;\delta^{+}(\omega^{2}
-{\vect{k}}^{2}-m^{2}) 
b_{\alpha}^{\dag}({\vect{k}};\omega)b_{\alpha}({\vect{k}};\omega)\nonumber\\
N_{d}(\omega) &=& \sum_{\alpha =1,2} \int
\frac{d^{3}{\vect{k}}}{(2\pi)^{3}}\;\delta^{+}(\omega^{2}
-{\vect{k}}^{2}-m^{2}) 
d_{\alpha}^{\dag}({\vect{k}};\omega)d_{\alpha}({\vect{k}};\omega).\nonumber 
\end{eqnarray}

\noindent Thus, the one-line replacements needed for $\frac{d\Delta \langle 
N_{b}(\omega) \rangle}{d\omega}$ are

\begin{eqnarray}
\langle  \psi_{\rho}(x){\overline{\psi}}_{\sigma}(y) 
\rangle_{\beta} &\rightarrow & \sum_{\alpha =1,2} \int 
\frac{d^{3}{\vect{k}}}{(2\pi)^{3}}\;\delta^{+}(\omega^{2}
-{\vect{k}}^{2}-m^{2}) \{
\langle \psi_{\rho}(x)b_{\alpha}^{ \dag}({\vect{k}};\omega) \rangle_{\beta} 
\langle b_{\alpha}({\vect{k}};\omega ) {\overline{\psi}}_{\sigma}(y) 
\rangle_{\beta}\nonumber\\
 && - \; \langle \psi_{\rho}(x)
b_{\alpha}({\vect{k}};\omega) \rangle_{\beta} \langle
b_{\alpha}^{\dag}({\vect{k}};\omega) {\overline{\psi}}_{\sigma}(y) 
\rangle_{\beta} \}\nonumber \\ &&\mbox{\vspace{5mm}}\nonumber\\
&=& \; \int \frac{d^{4}k}{(2\pi)^{3}}\; \delta^{+}(k^{2}-m^{2})\; \delta 
(k_{0}-\omega)\;(\not k + m)_{\rho \sigma} \nonumber\\
&& \; \times \;\{(1-f_{F}(\omega)) - 
f_{F}(\omega)(1 - f_{F}(\omega))\}e^{-ik(x-y)},\nonumber\\
\tselea{4.85}
\end{eqnarray}

\noindent where $f_{F}(\omega)$ is the Fermi-Dirac distribution: 
$f_{F}(\omega)=\frac{1}{e^{\beta (|\omega|-\mu)}+1}$, and  

\begin{eqnarray}
\langle {\cali{T}} ( \psi_{\rho}(x){\overline{\psi}}_{\sigma}(y))
\rangle_{\beta} &\rightarrow & \sum_{\alpha =1,2} \int
\frac{d^{3}{\vect{k}}}{(2\pi)^{3}}\;\delta^{+}(\omega^{2}
-{\vect{k}}^{2}-m^{2}) \;\{ \langle 
b_{\alpha}({\vect{k}};\omega)\psi_{\rho}(x) \rangle_{\beta} 
\langle b_{\alpha}^{\dag}({\vect{k}};\omega){\overline{\psi}}_{\sigma}(y) 
\rangle_{\beta} \nonumber\\ 
&& - \; \langle b_{\alpha}^{\dag}({\vect{k}};\omega) 
\psi_{\rho}(x) \rangle_{\beta} \langle 
b_{\alpha}({\vect{k}};\omega){\overline{\psi}}_{\sigma}(y) 
\rangle_{\beta}\}\nonumber \\ 
&&\mbox{\vspace{5mm}}\nonumber\\
&=& - \; \int \frac{d^{4}k}{(2\pi)^{3}}\; \delta^{+}(k^{2}-m^{2})\; \delta
(k_{0}-\omega)\;(\not k + m)_{\rho \sigma}\nonumber\\ 
&& \; \times \; f_{F}(\omega)(1 - f_{F}(\omega))e^{-ik(x-y)}.\nonumber\\
\tselea{4.9}
\end{eqnarray}

\noindent Correspondingly, for $\Delta \langle N_{b} \rangle$ we need

\begin{eqnarray}
\langle  \psi_{\rho}(x){\overline{\psi}}_{\sigma}(y)
\rangle_{\beta} &\rightarrow &  \int \frac{d^{4}k}{(2\pi)^{3}}\; 
\delta^{+}(k^{2}-m^{2})\;(\not k + m)_{\rho \sigma} \nonumber\\
&&\times \;\{(1-f_{F}(\omega)) -
f_{F}(\omega)(1 - f_{F}(\omega)\}e^{-ik(x-y)} \nonumber\\
&&\mbox{\vspace{1cm}}\nonumber\\
\langle {\cali{T}} ( \psi_{\rho}(x){\overline{\psi}}_{\sigma}(y))
\rangle_{\beta} &\rightarrow& - \; \int \frac{d^{4}k}{(2\pi)^{3}}\; 
\delta^{+}(k^{2}-m^{2})\;(\not k + m)_{\rho \sigma}
f_{F}(\omega)(1 - f_{F}(\omega))e^{-ik(x-y)}.\nonumber\\
\tselea{4.10}
\end{eqnarray}

\noindent For the $d$-type excitations the prescription is very similar, 
actually, in order to get $\frac{d\Delta \langle N_{d}(\omega)
\rangle}{d\omega}$, the following substitutions must be performed in
(\tseref{4.85})-(\tseref{4.10}): $\theta(k_{0}) \rightarrow
\theta(-k_{0})$, $f_{F}\rightarrow (1-f_{F})$ and $\mu \rightarrow -\mu$.

\vspace{1cm}

\noindent {\bf (iii) For gauge fields in the axial temporal gauge 
($A^{0}=0$):}

\vspace{4mm}

\noindent The temporal gauge is generally incorporated in the gauge fixing
sector of the Lagrangian and particularly

\begin{equation}
{\cali{L}}_{fix}=-\frac{1}{2\alpha}(A_{0})^{2}; \alpha
\rightarrow 0. \tseleq{gauge fixing}
\end{equation}

\noindent The principal advantage of the axial gauges arises from the
decoupling the F-P ghosts in the theory. This statement is of course
trivial in QED as any linear gauge (both for covariant and noncovariant
case) brings this decoupling automatically \tsecite{LB}. Particular
advantage of the temporal gauge comes from an elimination of non-physical 
scalar photons from the very beginning. 

\vspace{3mm}

\noindent Let us decompose a gauge field $A_{i}, i=1,2,3$ into the 
transverse and longitudial part, i.e. $A_{i}=A_{i}^{T}+A_{i}^{L}$ with

\begin{equation}
A_{i}^{T}=\left( \delta_{ij} - \frac{\partial_{i}
\partial_{j}}{{\vec{\partial}}^{2}} \right) A_{j} \;\; \mbox{and}\;\; 
A_{i}^{L}=\frac{\partial_{i}
\partial_{j}}{{\vec{\partial}}^{2}} A_{j}, \nonumber
\end{equation}

\vspace{3mm}
\noindent and use the sum over gauge-particle polarizations 

\begin{equation}
\sum_{\lambda = 1}^{2} {\varepsilon}^{(\lambda)}_{i}(k)
{\varepsilon}^{(\lambda)}_{j}(k) = {\delta}_{ij} -
\frac{k_{i}k_{j}}{{\vect{k}}^{2}},\nonumber
\end{equation}

\noindent with ${\varepsilon}^{(\lambda)}(k)$ being polarization vectors, 
then for $\frac{d\Delta \langle N^{T}(\omega) \rangle}{d\omega}$ we get the 
following one-line replacements 

\begin{eqnarray}
\langle A^{T}_{i}(x)A^{T}_{j}(y) {\rangle}_{\beta} &\rightarrow& 
\sum_{\lambda = 1}^{2}
\int
\frac{d^{3}{\vect{k}}}{(2\pi)^{3}}\;\delta^{+}(\omega^{2}  
-{\vect{k}}^{2}-m^{2}) \langle 
A^{T}_{i}(x)a^{\dag}_{\lambda}({\vect{k}};\omega)
{\rangle}_{\beta} \langle a_{\lambda}({\vect{k}};\omega)A^{T}_{j}(y) 
{\rangle}_{\beta}\nonumber \\
&&+ \;\langle A^{T}_{i}(x)a_{\lambda}({\vect{k}};\omega) 
{\rangle}_{\beta} \langle
a^{\dag}_{\lambda}({\vect{k}};\omega)A^{T}_{j}(y) {\rangle}_{\beta} 
\}\nonumber\\ &&\mbox{\vspace{5mm}}\nonumber\\
&=&\left( \delta_{ij} - \frac{\partial_{i}
\partial_{j}}{{\vec{\partial}}^{2}} \right) 
(\mbox{Eq.(\tseref{new1})})
\nonumber\\ &&\mbox{\vspace{1.5cm}}\nonumber\\
\langle {\cali{T}} (A^{T}_{i}(x)A^{T}_{j}(y) ) {\rangle}_{\beta} 
&\rightarrow &
\sum_{\lambda = 1}^{2} \int
\frac{d^{3}{\vect{k}}}{(2\pi)^{3}}\;\delta^{+}(\omega^{2}
-{\vect{k}}^{2}-m^{2}) \{ \langle 
A^{T}_{i}(x)a^{\dag}_{\lambda}({\vect{k}};\omega){\rangle}_{\beta}
\langle A^{T}_{i}(y)a_{\lambda}({\vect{k}};\omega) 
{\rangle}_{\beta}\nonumber \\ &&+\; \langle 
A^{T}_{i}(x)a_{\lambda}({\vect{k}};\omega){\rangle}_{\beta}\langle
A^{T}_{i}(x)a^{\dag}_{\lambda}({\vect{k}};\omega) {\rangle}_{\beta}\nonumber\\
&&\mbox{\vspace{5mm}}\nonumber\\
&=&\left( \delta_{ij} - \frac{\partial_{i}
\partial_{j}}{{\vec{\partial}}^{2}} \right) (\mbox{Eq.(\tseref{new4})}).
\tselea{4.11}
\end{eqnarray}

\noindent The replacements needed for $\Delta \langle N^{T} \rangle$ can 
be concisely expressed as

\begin{eqnarray}
\langle \ldots \rangle_{\beta} \rightarrow \left( \delta_{ij} - 
\frac{\partial_{i}
\partial_{j}}{{\vec{\partial}}^{2}}\right) (\mbox{Eq.(\tseref{new2})})
\tselea{4.12}
\end{eqnarray}

\noindent As for the longitudial (non-physical) degrees of freedom, it is 
obvious that

\begin{equation}
\langle A^{L}_{i}(x)A^{L}_{j}(y){\rangle}_{\beta}; 
 \; \;\langle 
{\cali{T}} (A^{L}_{i}(x) 
A^{L}_{j}(y) )  {\rangle}_{\beta} \rightarrow 0. \end{equation}

\vspace{4mm}

\noindent Eqs.(\tseref{new1})-(\tseref{4.12}) can be most easily derived in
the finite volume limit, e.g. for a scalar field we reformulate
$\varphi(x)$ as

\begin{equation}
\varphi(x) = \sum_{r} \frac{A_{r}}{\sqrt{2E_{r}V}}e^{-iE_{r}t +
i{\vect{k}}_{r}{\vect{x}}} +
\frac{A^{\dag}_{r}}{\sqrt{2E_{r}V}}e^{iE_{r}t-i{\vect{k}}_{r}{\vect{x}}},
\nonumber \end{equation}

\noindent rescaling the annihilation and creation operators by defining
$a(k) = \sqrt{2E_{k}V} A_{k}$ in such a way that $[A_{k};A^{\dag}_{k^{'}}] =
\delta_{kk^{'}}$ (so that $ {\langle A^{\dag}_{k}A_{k^{'}}
\rangle}_{\beta} = \delta_{kk^{'}}f_{B}(k_{0})$), while $\int  
\frac{d^{3}{\vect{k}}}{(2\pi)^{3}} \rightarrow \frac{1}{V}\sum_{\vect{k}}$.

\vspace{3mm}

\begin{figure}[h] \vspace{3mm} \epsfxsize=11.5cm
\centerline{\epsffile{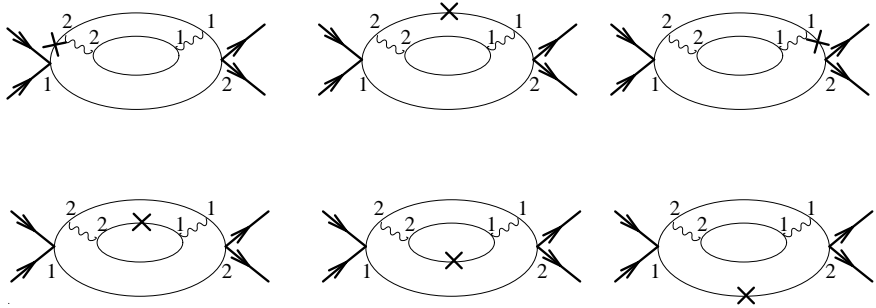}} \caption{{\it The numerator of
(\tseref{dN}) and (\tseref{N}) can be calculated using the modified cut
diagrams for $\langle T^{\dag} {\cali{P}} T \rangle_{p_{1} p_{2}, \beta}$.
As an example we depict all the possible contributions to the numerator
derived from the cut diagram on Fig.\ref{fig.3} c). The wavy lines and
thin lines describe the heat-bath particles. The crossed lines denote the
substituted propagators, in this case we wish to calculate the thin-line
particle number spectrum.}} \label{fig7} \vspace{1mm} \end{figure}

\noindent The replacements (\tseref{new1})-(\tseref{4.12}) are meant in
the following sense: firstly one constructs all the $T \not= 0$ diagrams
for ${}\langle T^{\dag}{\cali{P}}T \rangle_{p_{1}p_{2}, \beta} $, using
the LTE (\tseref{proj4}) and the rules mentioned therein. In order to
calculate the numerator of (\tseref{dN}) or (\tseref{N}) we simply replace
(using corresponding prescriptions) $one$ heat bath particle line in each
cut diagram and this replacement must sum for all the possible heat-bath
particle lines in the diagram. If more types of heat-bath particles are
present, we replace only those lines which correspond to particles whose
number spectrum we want to compute (see Fig.\ref{fig7}). 

\vspace{2mm}

\noindent The terms in the replacements (\tseref{new1})-(\tseref{4.12})
have a direct physical interpretation. The $f(\omega_{k})$ and
$(1+(-1)^{p}f(\omega_{k}))$ can be viewied as the absorption and emission
of the heat-bath particles respectively \tsecite{PVLJT}. The term
$f(\omega_{k})(1+(-1)^{p}f(\omega_{k}))$ describes the fluctuations of the
heat bath particles. This is because for the non-interacting heat-bath
particles $\langle (n_{k}-\langle n_{k}\rangle_{\beta})^{2}
\rangle_{\beta} = f(\omega_{k})(1+(-1)^{p}f(\omega_{k}))$. The substituted
propagators can be therefore schematically depicted as

\vspace{1mm}
\begin{figure}[h]
\hspace{-2cm}  
\epsfxsize=4cm
\centerline{\epsffile{fig16.eps}}
\label{fig16}
\end{figure}
\vspace{2mm}

\noindent Collecting all the contributions from emissions, absorptions and 
fluctuations separately, one can schematically write

\begin{equation}
\frac{d\langle N(\omega)\rangle_{f}}{d\omega}=\frac{d\langle 
N(\omega)\rangle_{i}}{d\omega} + F^{emission}(\omega) + 
F^{absorption}(\omega)+F^{fluc}(\omega), 
\tseleq{L-T}
\end{equation}

\vspace{1mm}
\noindent where, for instance for neutral scalar bosons

\vspace{-3mm}

\begin{eqnarray*}
F^{emission}(\omega)= Z^{-1}_{f}\int \frac{d^{4}k}{(2\pi)^{3}} 
\delta^{+}(k^{2}-m^{2})\delta(k_{0}-\omega)(1+f_{B}(\omega))\left\langle
\frac{T^{\dag} {\stackrel{\leftarrow}{\delta}}}{\delta 
\psi(x)}{\cali{P}}
\frac{{\stackrel{\rightarrow}{\delta}} T}{\delta \psi(y)}
\right\rangle_{p_{1} p_{2}, \beta}. 
\end{eqnarray*}

\noindent Using (\tseref{new4}), it is easy to write down the analogous
expressions for the $F^{absorption}$ and $F^{fluc}$. To the lowest
perturbative order, the form (\tseref{L-T}) was obtained by Landshoff and
Taylor \tsecite{PVLJT}.

\section{Model process}
\subsection{Basic assumptions}

To illustrate the modified cut diagram technique, we shall restrict
ourselves to a toy model, namely to a scattering of two neutral scalar
particles $\Phi$ within a photon-electron heat bath, with a pair of scalar
charged particles {$\phi, {\overline\phi}$} left as a final product. Both
initial and final particles are supposed to be unheated. We further assume
that the heat bath photons $A$ and electrons $\Psi$ are scalars, i.e. the
heat-bath Hamiltonian has form

\vspace{-2mm}
\begin{eqnarray} H^{hb} &=& H^{\gamma} + H^{e} + eA\Psi {\Psi}^{\dag}
+\frac{e^{2}}{2}A^{2} \Psi {\Psi}^{\dag} \nonumber \\ H^{e} &=&
\partial_{\nu} \Psi \partial^{\nu} {\Psi}^{\dag} - m_{e}^{2} \Psi
{\Psi}^{\dag}\nonumber \\ H^{\gamma} &=& \frac{1}{2} (\partial_{\nu}A)^{2}
- \frac{m_{\gamma}^{2}}{2}A^{2}. \vspace{2mm} \tselea{Hamilt} \end{eqnarray}

\noindent On the other hand $H_{in}$ entering in the $T$ - matrix reads
$H_{in} = \frac{\lambda}{2}\Phi^{2}\phi {\phi}^{\dag} + (eA +
\frac{e^{2}}{2}A^{2})\Psi {\Psi}^{\dag} + (eA + \frac{e^{2}}{2}A^{2})\phi
{\phi}^{\dag}$. It is usually argued \tsecite{NS, EP} that the interacting
pieces in $H^{hb}$ can be dropped provided that $t_{i} \rightarrow
-\infty$ and $t_{f} \rightarrow \infty$. In the following we accept this
omission as it allows us to use safely Wick's theorem (\tseref{wick}) and
Dyson-Schwinger equation (\tseref{S-D2}).

\subsection{Calculations}

\begin{figure}[h]
\epsfxsize=11cm
\centerline{\epsffile{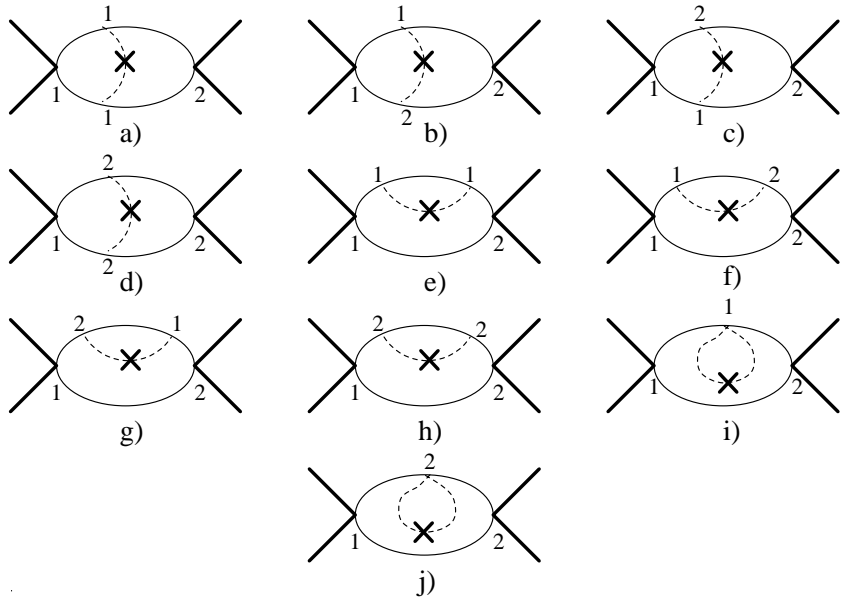}}
\caption{\em The modified cut diagrams involved in an order-$e^{2}$ contribution 
to the photon number spectrum. Dashed lines: photons. Solid lines: $\phi$, 
$\phi^{\dag}$ particles. Bold lines: $\Phi$ particles.} 
\label{fig35}
\end{figure}

We can now compute an order-$e^{2}$ contribution to the $\frac{d\Delta
\langle N_{\gamma}(\omega) \rangle}{d\omega}$. The evaluation of the
$\frac{d\Delta \langle N_{\gamma}(\omega) \rangle}{d\omega}$ is
straightforward. In Fig.\ref{fig35} we list all the modified cut diagrams
contributing to an order-$e^{2}$.

\vspace{2mm}
\vspace{2mm}

\noindent Note that diagrams b) and c) are topologically identical. 
Similarly, diagrams e), f), h), i) and j) should be taken with
combinatorial factor 2 (corresponding diagrams for insertion of a
heat-bath particle line in the bottom, solid line are omited). Of course,
diagram g) vanishes for kinematic reasons. Let us emphasize that it is
necessary to give sense to graphs e), h), i) and j) as these suffer with
the pinch singularity; the muon-particle propagator
$(p_{1;2}^{2}-m^{2})^{-1}$ has to be evaluated at its pole because of the
presence of an on-shell line (1-2 line) with the same momenta. Some
regularization is obviously necessary. Using the formal identity
\tsecite{LB}

\begin{equation}
\frac{1}{x+i\varepsilon}\delta(x)= 
-\frac{1}{2}\delta^{'}(x)-i\pi(\delta(x))^{2},
\end{equation}

\vspace{3mm}

\noindent we discover that the unwanted $\delta^{2}$ mutually cancel
between e) and h) diagrams (similarly for i) and j) diagrams). An 
alternative (but lengthier)) way of dealing with the latter pinch 
singularity; i.e. switching off the interaction with a heat bath in 
the remote past and future, is discussed in \tsecite{PVLJ}. Using the
prescriptions (\tseref{new1})-(\tseref{new4}) and evaluating the 
integrals
(note, we should attach to each digram the factor $\frac{1}{2!}$ coming
from a Taylor expansion of the $T$-matrix), we are left with (c.f.
Eq.(\tseref{L-T})): 
 
\begin{eqnarray}
\lefteqn{F^{emission}(\omega) + F^{absorption}(\omega)}\nonumber\\
&=&\frac{t\lambda^{2}e^{2}}{\langle 
T{\cali{P}}T^{\dag}\rangle_{p_{1}p_{2}, \beta}V 8 
\omega_{p_{1}}\omega_{p_{2}}(2\pi)^{5}}\int d^{4}k \;
\delta(k^{2}-m_{\gamma}^{2}) \;
\delta(k_{0}-\omega)\nonumber\\
&&\times \int  d^{4}q_{1}d^{4}q_{2} \; \delta^{+}(q_{1}^{2}-m_{\mu}^{2})
\; \delta^{+}(q_{2}^{2}-m_{\mu}^{2})\nonumber\\
&&\times ~\left\{ 
K_{1}(1+f(\omega)) \; \delta^{4}(-Q+q_{1}+q_{2}+k)\right. \nonumber\\
&&\left.-K_{2}f_{B}(\omega)
\; \delta^{4}(-Q+q_{1}+q_{2}-k)\right\}
\tselea{30}
\end{eqnarray}

and

\begin{eqnarray}
F^{fluct}(\omega)&=&\frac{t\lambda^{2}e^{2}f_{B}(\omega)(1+f_{B}(\omega))}{\langle
T{\cali{P}}T^{\dag}\rangle_{p_{1}p_{2}, \beta}V 8
\omega_{p_{1}}\omega_{p_{2}}(2\pi)^{5}}\int d^{4}k 
\; \delta(k^{2}-m_{\gamma}^{2})\; \delta(k_{0}-\omega)\nonumber\\
&&\times \int  d^{4}q_{1}d^{4}q_{2} \; \delta^{+}(q_{1}^{2}-m_{\mu}^{2})\;
\delta^{+}(q_{2}^{2}-m_{\mu}^{2})\nonumber\\
&&\times \left\{ 
\delta^{4}(-Q+q_{1}+q_{2}+k)K_{1}+\delta^{4}(-Q+q_{1}+q_{2}-k)K_{2}\right.\nonumber\\
&&\left.-~2\delta^{4}(-Q+q_{1}+q_{2})K_{3} \right\}\nonumber\\
&&\mbox{\vspace{5mm}}\nonumber\\
&+&~\frac{t\lambda^{2}e^{2}f_{B}(\omega)(1+f_{B}(\omega))}{\langle
T{\cali{P}}T^{\dag}\rangle_{p_{1}p_{2}, \beta}V 8 
\omega_{p_{1}}\omega_{p_{2}}(2\pi)^{5}}\int d^{4}k
\; \delta(k^{2}-m_{\gamma}^{2}) \; \delta(k_{0}-\omega)\nonumber\\
&&\times \int 
d^{4}q_{1}d^{4}q_{2} \; \delta^{4}(-Q+q_{1}+q_{2})\nonumber\\
&&\times \left\{ 
\left(1-\frac{1}{2q_{1}k 
-m^{2}_{\gamma}}+ 
\frac{1}{2q_{1}k + 
m^{2}_{\gamma}}\right)\delta^{+}(q_{2}^{2}-m_{\mu}^{2})\frac{\partial}{\partial 
m^{2}_{\mu}}\delta^{+}(q_{1}^{2}-m_{\mu}^{2})\right. \nonumber\\
&&\left. + (q_{1} \leftrightarrow 
q_{2})\right\} 
\tselea{31}
\end{eqnarray}

\vspace{3mm}

\noindent with $K_{1}=\left( \frac{1}{2q_{1}k + m^{2}_{\gamma}}+
\frac{1}{2q_{2}k + m^{2}_{\gamma}}\right)^{2}$, $K_{2}=\left(
\frac{1}{2q_{1}k - m^{2}_{\gamma}}+ \frac{1}{2q_{2}k -
m^{2}_{\gamma}}\right)^{2}$,
$K_{3}=\frac{2}{(2q_{1}k-m^{2}_{\gamma})(2q_{2}k+m^{2}_{\gamma})}$ and
$Q=p_{1}+ p_{2}$. We have dropped the $i\varepsilon$ prescription in the
propagators since adding/ subtracting an on-shell momenta $q_{1;2}$ to/from
an on-shell momenta $k$ we can not fulfil the condition $(k\pm
q_{1;2})^{2}=m^{2}_{\mu}$. As it is usual, we have assumed that our
interaction is enclosed in a `time' and volume box ($t$ and $V$
respectively). The relevant (i.e. order-$e^{0}$) term for $\langle T
{\cali{P}}T^{\dag}\rangle_{p_{1}p_{2}, \beta}$ reads

\begin{eqnarray}
\langle T {\cali{P}}T^{\dag}\rangle_{p_{1}p_{2}, \beta}&=&
\frac{\lambda^{2}t}{8V\omega_{p_{1}}\omega_{p_{2}}(2\pi)^{2}}\int 
d^{4}q_{1}d^{4}q_{2} \; \delta^{+}(q_{1}^{2}-m_{\mu})\; 
\delta^{+}(q_{2}^{2}-m_{\mu}^{2})\; \delta^{4}(-Q+q_{1}+q_{2})\nonumber\\
&=& \frac{\lambda^{2}t}{64V\omega_{p_{1}}\omega_{p_{2}}|Q| 
\pi}\sqrt{Q^{2}-4m_{\mu}^{2}},
\end{eqnarray}

\vspace{3mm}

\noindent Eqs.(\tseref{30}) and (\tseref{31}) are 
analogous to the result obtained in \tsecite{PVLJT} for the decay. 

\vspace{3mm}

\begin{figure}[h] \epsfxsize=11cm
\centerline{\epsffile{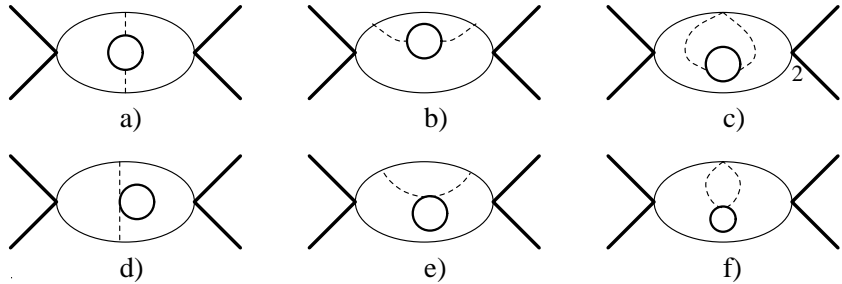}} \caption{\em The generating thermal
diagrams involved in an order-$e^{2}$ contribution to the photon number
spectrum. Dashed lines: photons. Thin lines: $\phi$, $\phi^{\dag}$
particles. Bold lines: $\Phi$ particles. Half-bold lines:  electrons.}
\label{fig55} \end{figure}

\noindent One can perform the similar calculations for the electron number
spectrum $\frac{d\Delta \langle N_{e}(\omega) \rangle}{d\omega}$. The task
is now, however, tougher. The major difference in comparison with
the photon number spectrum is that the lowest order in $e$ (keeping
$\lambda^{2}$) is $e^{4}$. This brings richer diagrammatic structure
then in the photon case. In Fig.\ref{fig55} we list all the generating
thermal diagrams contributing to an order-$e^{4}$.

\vspace{2mm}  \vspace{2mm}
 
\noindent It is easy to see that out of these 6 generating thermal
diagrams we get 41 non-vanishing cut diagrams (10 from a); 9 from b); 4
from c); 8 from d); 4 from e) and 6 from f).) The actual electron number
spectrum calculations are thus rather long and not extremely rewarding for our
unphysical toy model, so we refrain from doing them.

\section{Conclusions}

In this paper we have formulated a systematic method for studying the
heat-bath particle number spectrum using modified cut diagrams. In
particular, for the quark-gluon plasma in thermodynamical equilibrium our
approach should be useful as an effective alternative to the Landshoff and
Taylor \tsecite{PVLJT} approach. The method used in \tsecite{PVLJT} (i.e. 
to start from first principles) suffers from the lack of a systematic
computational approach for higher orders in coupling constants. One of the
corner stones of our formalism is the largest-time equation (LTE). We have
shown how the zero-temperature LTE can be extended to finite temperature. 
During the course of this analysis, we have emphasized some important
properties of the finite-temperature extension which are worth mentioning.
Firstly, many of kinematic rules valid for zero-temperature diagrams can
not be directly used in the finite-temperature ones. This is because the
emission or absorption of heat-bath particles make it impossible to fix
some particular direction to a diagrammatic line. It turns out that one
finds more diagrams then one used to have at zero temperature. The most
important reductions of the diagrams have been proved. The rather
complicated structure of the finite-temperature diagrams brings into play
another complication: uncutable diagrams. It is well known that at zero 
temperature one can always make only one cut in each cut diagram (this can be
viewed as a consequence of the unitarity condition). This is not true
however at finite temperature. We have found it as useful to start fully
with the LTE analysis which is in terms of type 1 and type 2 vertices.
This language allows us to construct systematically all the cut diagrams.
We have refrained from an explicit use of the cuts in finite-temperature
diagrams as those are ambiguous and therefore rather obscure the 
analysis. 

\vspace{3mm}

\noindent The second, rather technical, corner stone is the thermal
Dyson-Schwinger equation. We have developed a formalism of the {\em
arrowed} variations acting directly on field operators. This provides an
elegant technique for dealing in a practical fashion with expectation
values (both thermal and vacuum) whenever functions or functionals of
fields admit the decomposition (\tseref{Hin2}). The merit of the
Dyson-Schwinger equation is that it allows us to rewrite an expectation
value of some functional of field in terms of expectation values of less
complicated functionals. Some illustrations of this and further thermal
functional identities are derived in Appendix A. 

\vspace{3mm}

\noindent When we have studied the heat-bath particle number spectrum, we
applied the Dyson-Schwinger equation both to numerator and denominator of
corresponding expression. The results were almost the same. The simple
modification of one propagator rendered both equal. We could reflect this
on a diagrammatical level very easily as the denominator was fully
expressible in terms of thermal cut diagrams. Our final rule for the
heat-bath particle spectrum is

\begin{equation}
\frac{d \Delta \langle N(\omega) \rangle}{d\omega} = \frac{\langle 
T^{\dag}{\cali{P}}T \rangle_{p_{1}p_{2},\beta}^{M}}{\langle 
T^{\dag}{\cali{P}}T \rangle_{p_{1}p_{2},\beta}},\nonumber
\end{equation}

\noindent with $T$ being the $T$-matrix, ${\cali{P}}$ being the projection
operator onto final states, $p_{1}, p_{2}$ being the momenta of particles
in the initial state, $\beta$ being the inverse temperature and $M$ being
abbreviation for the modified digrams. Modification of the cut diagrams
consist of the substitution in turn of each heat-bath particle line by an
altered one. This substitution must be done in each cut diagram. 
Replacement must be only one per modified diagram. Our approach is
demonstrated on a simple model where two scalar particles (``pions'')
scatter, within a photon-electron heat bath, into a pair of charged
particles (``muon'' and ``antimuon'') and it is shown how to calculate the
resulting changes in the number spectra of the photons and electrons.

\section*{Acknowledgements}

We are indebted to P.V.Landshoff for reading the manuscript and for
invaluable discussion. The work is supported in part by Fitzwilliam
College scholarship.

\appendix
\section{Appendix}
\subsection{Functional formalism: general background}

Eq.(\tseref{S-D2}) gives us an alternative definition of the
Dyson-Schwinger equation in terms of the ``functional derivation" 
$\frac{\delta}{\delta \psi(x)}$. Let us first show that (\tseref{S-D2}) is
consistent with Wick's theorem (\tseref{wick})-(\tseref{wick2}). To be
specific, let us consider an ensemble of non-interacting particles in
thermodynamical equilibrium. In order to keep the work transparent, we
shall suppress all the internal indices. There is no difficulty whatsoever
in reintroducing the necessary details.  Let us first realize that for any
(well behaved) functional the following Taylor's expansion holds
\tsecite{PR}

\begin{equation} X[\psi] = \sum_{n} \int dx_{1} \ldots \int dx_{n}
\alpha^{n}(x_{1}\ldots x_{n}) \psi(x_{1})\ldots \psi(x_{n}), 
\tseleq{Hin2}
\vspace{1mm} \end{equation}

\noindent The same is true if $\psi$ is an operator instead. In the latter
case the $\alpha^{n}(\ldots)$ are not generally symmetric in the $x$'s
{\footnote{If $X = X[\psi, \partial\psi]$, the $\alpha^{n}$ may also
contain derivations working on the various fields. In this paper we rule
out such a case from our reasonings.}}. When Fermi fields are involved, we
might, for the sake of compactness, include in the argument of $\psi$ the
space-time coordinate, the Dirac index, and a discrete index which
distinguishes $\psi_{\alpha}$ from ${\overline{\psi}}_{\alpha}$. In the
latter case $\int dx \rightarrow \sum \int dx$, whera summation runs over
the discrete indices. With this convention, the expansion (\tseref{Hin2})
holds even for the Fermi fields.  An extension of (\tseref{Hin2}) to the
case where different fields are present is natural. Particularly important
is the case when $\psi$ is a field in the interaction picture, using
Wick's theorem and decomposition (\tseref{Hin2}) one can then write

\begin{eqnarray}
\lefteqn{\langle G[\psi] \psi(x) F[\psi] \rangle_{\beta} =}\nonumber\\ 
&=&\sum_{m,n} \left(\int dx \right)^{n}
\left( \int dy \right)^{m} \alpha^{n}(x_{1} \ldots x_{n}) \beta^{m}(y_{1} 
\ldots  y_{m})
\left\langle \left(\prod_{k}^{n}\psi(x_{k})\right) \psi(x) 
\prod_{k'}^{m}\psi(y_{k'}) \right\rangle_{\beta}\nonumber\\ 
&=& \sum_{n} \left(\int dx \right)^{n} \alpha^{n}(x_{1} 
\ldots x_{n})  \sum_{l}^{n} (\pm 1)^{n-l} \langle 
\psi(x_{l}) \psi(x) \rangle_{\beta} \left\langle \prod_{k \not= l}^{n} 
\psi(x_{k}) 
F[\psi] \right\rangle_{\beta}  \nonumber\\ 
&+&  \sum_{m} \left(\int dy
\right)^{m} \beta^{m}(y_{1} \ldots y_{m}) \sum_{l}^{m} (\pm 1)^{l-1} \langle 
\psi(x) \psi(x_{l})\rangle_{\beta} \left\langle 
G[\psi] \prod_{k' \not= l}^{m} \psi(y_{k'} ) 
\right\rangle_{\beta}.
\tselea{D-S3}
\end{eqnarray}

\noindent with $\left( \int dx \right)^{n}=\int dx_{1} \ldots \int dx_{n}$. 
The `$-$' stands for fermions and `$+$' for bosons. On the other hand, using 
the formal prescriptions (\tseref{var1}) and
(\tseref{var2}) for  $\frac{{\stackrel{\rightarrow}{\delta}}}{\delta 
\psi(x)}$ one can read

\begin{eqnarray}
\lefteqn{\int dz \langle \psi(x) \psi(z) \rangle_{\beta} \left\langle G[\psi]
\frac{{\stackrel{\rightarrow}{\delta}} F[\psi]}{\delta \psi (z)} 
\right\rangle_{\beta} =}\nonumber\\ 
&=&\mbox{\hspace{5mm}} \sum_{m} \left( \int dy \right)^{m} \beta^{m}(y_{1} 
\ldots y_{m})
\int dz \langle \psi(x) \psi(z) \rangle_{\beta} \sum_{l}^{m} (\pm 1)^{l-1} 
\delta(z-y_{l}) \left\langle G[\psi] \prod_{k' \not= l}^{m} \psi(y_{k'}) 
\right\rangle_{\beta}\nonumber\\
&=&\mbox{\hspace{5mm}} \sum_{m} \left( \int dy \right)^{m} \beta^{m}(y_{1} 
\ldots y_{m}) \sum_{l}^{m} (\pm 1)^{l-1}\langle \psi(x) \psi(y_{l}) 
\rangle_{\beta} \left\langle G[\psi]
\prod_{k' \not= l}^{m} \psi(y_{k'}) \right\rangle_{\beta}.
\end{eqnarray}

\noindent Similar expression holds for $\int dz \langle \psi(x) \psi(z)
\rangle_{\beta} \left\langle \frac{ G[\psi]
{\stackrel{\leftarrow}{\delta}}}{\delta \psi (z)} F[\psi]
\right\rangle_{\beta}$. Putting latter two together we get precisely
(\tseref{D-S3}). This confirms the validity of (\tseref{S-D2}). It is easy
to persuade oneself that exactly the same sort of arguments leads to

\begin{eqnarray}
&&\langle \psi(x) F[\psi] \rangle_{\beta} = \int dz \langle \psi(x) \psi(z) 
\rangle_{\beta} \left\langle \frac{{\stackrel{\rightarrow}{\delta}} 
F[\psi]}{\delta \psi(z)} \right\rangle_{\beta}\\
\tselea{S-D3}
&&\langle {\cali{T}}(\psi(x) F[\psi]) \rangle_{\beta} =\int dz \langle 
{\cali{T}}(\psi(x) \psi(z)) \rangle_{\beta} \left\langle {\cali{T}} 
\left( \frac{{\stackrel{\rightarrow}{\delta}} F[\psi]}{\delta \psi(z)} 
\right) \right\rangle_{\beta}\\ \tselea{S-D4}
&&\langle G[\psi] {\cali{T}}( \psi(x) F[\psi]) \rangle_{\beta} = \int dz 
\langle {\cali{T}}(\psi(x) \psi(z)) \rangle_{\beta} \left\langle 
G[\psi] {\cali{T}} \left( \frac{{\stackrel{\rightarrow}{\delta}} 
F[\psi]}{\delta \psi(z)} \right) \right\rangle_{\beta} +  \nonumber\\
&& \mbox{\hspace{3.7cm}} + \int dz \langle \psi(z) \psi(x) 
\rangle_{\beta} 
\left\langle \frac{ G[\psi] {\stackrel{\leftarrow}{\delta}}}{\delta 
\psi(z)} {\cali{T}}(F[\psi]) \right\rangle_{\beta},\\
&&\mbox{etc.}\nonumber
\tselea{S-D6}
\end{eqnarray}

\noindent with ${\cali{T}}$ being either the chronological or
anti-chronological time ordering symbol. At this stage it is important to
realize that from the definition of
$\frac{{\stackrel{\rightarrow}{\delta}}}{\delta \psi(x)}$ directly follows
that $[\frac{{\stackrel{\rightarrow}{\delta}}}{\delta \psi(x)}; 
\frac{{\stackrel{\rightarrow}{\delta}}}{\delta \psi(y)}]_{\mp}=0$ ( `$-$'
holds for bosons and `$+$' for fermions). Indeed,

\vspace{-2mm}
\begin{eqnarray}
\frac{{\stackrel{\rightarrow}{\delta^{2}}}F[\psi]}{\delta \psi(x) \delta 
\psi(y)} &=& \sum_{n=2} \sum_{i < j} \left(\int dx \right)^{n-2} 
(\alpha^{n}(x_{1} 
\ldots \stackrel{x_{i}}{\stackrel{\downarrow}{x}} \ldots 
\stackrel{x_{j}}{\stackrel{\downarrow}{y}} \ldots x_{n})  \pm \nonumber\\ 
&\pm &~\alpha^{n}(x_{1} \ldots
\stackrel{x_{i}}{\stackrel{\downarrow}{y}} \ldots 
\stackrel{x_{j}}{\stackrel{\downarrow}{x}} \ldots 
x_{n}))(\pm1)^{i+j} \prod_{m \not= 
i,j}^{n} \psi(x_{m}) = \mp 
\frac{{\stackrel{\rightarrow}{\delta^{2}}}F[\psi]}{\delta 
\psi(y) \delta \psi(x)}.\nonumber\\
\tselea{commut5}
\end{eqnarray}

\noindent Similarly $[\frac{{\stackrel{\leftarrow}{\delta}}}{\delta
\psi(x)};  \frac{{\stackrel{\leftarrow}{\delta}}}{\delta
\psi(y)}]_{\mp}=0$. Analogously we might prove

\begin{equation}
\frac{F[\psi]{\stackrel{\leftarrow}{\delta^{2}}}}{\delta \psi(x)\delta 
\psi(y)}=
\frac{{\stackrel{\rightarrow}{\delta^{2}}}F[\psi]}{\delta \psi(x)\delta 
\psi(y)} \end{equation}

\noindent and

\begin{eqnarray}
\frac{{\stackrel{\rightarrow}{\delta^{2}}}(F[\psi]G[\psi])}{\delta 
\psi(x)\delta 
\psi(y)} &=& \frac{F[\psi]{\stackrel{\leftarrow}{\delta^{2}}}}{\delta 
\psi(x)\delta \psi(y)}G[\psi] + (-1)^{p}
\frac{F[\psi]{\stackrel{\leftarrow}{\delta}}}{\delta \psi(x)}
\frac{{\stackrel{\rightarrow}{\delta}}G[\psi]}{\delta \psi(y)}\nonumber\\
&+& \frac{F[\psi]{\stackrel{\leftarrow}{\delta}}}{\delta \psi(y)}
\frac{{\stackrel{\rightarrow}{\delta}}G[\psi]}{\delta \psi(x)}
+F[\psi]\frac{{\stackrel{\rightarrow}{\delta^{2}}}G[\psi]}{\delta 
\psi(x)\delta \psi(y)}.
\end{eqnarray}

\noindent The ``$p$'' is $0$ for bosons and $1$ for fermions. With
(\tseref{S-D2}) and (A.4) - (A.6) one can easily construct more
complicated expectation values. For example, using (\tseref{S-D2}) and
(A.4) we get

\begin{eqnarray}
\lefteqn{\langle \psi(x) \psi(y) F[\psi] \rangle_{\beta} =}\nonumber\\
&=& \int 
\frac{dz_{1}dz_{2}}{2} (\langle \psi(x) \psi (z_{1}) \rangle_{\beta} 
\langle \psi(y) \psi (z_{2})  \rangle_{\beta} + (-1)^{p}\langle \psi(x) 
\psi(z_{2}) \rangle_{\beta} \langle \psi(y) \psi(z_{1}) 
\rangle_{\beta})\nonumber\\ 
&& \times \; \left\langle \frac{{\stackrel{\rightarrow}{\delta^{2}}} 
F[\psi]}{\delta 
\psi(z_{1}) \delta \psi(z_{2})} \right\rangle_{\beta}\nonumber\\
&+& \langle  \psi(x) \psi(y) \rangle_{\beta} \langle F[\psi]\rangle_{\beta}.
\tselea{S-D7}
\end{eqnarray}

\noindent Similarly, using (\tseref{S-D2}) and (anti-)commutativity of the 
arrowed $\frac{\delta}{\delta \psi(x)}$, we get

\begin{eqnarray} 
\lefteqn{\langle G[\psi]  \psi(x) \psi(y) F[\psi] \rangle_{\beta} 
=}\nonumber\\
&=& \int \frac{dz_{1} dz_{2}}{2} (\langle \psi(x) 
\psi(z_{1})\rangle_{\beta} \langle \psi(y) \psi(z_{2})\rangle_{\beta} + 
(-1)^{p}\langle \psi(x) 
\psi(z_{2}) \rangle_{\beta} \langle \psi(y) \psi(z_{1}) 
\rangle_{\beta})\nonumber\\
&&\times ~ \left\langle G[\psi] 
\frac{{\stackrel{\rightarrow}{\delta^{2}}}F[\psi]}{\delta  
\psi(z_{1}) \delta \psi(z_{2})} \right\rangle_{\beta}\nonumber\\
&+& \int \frac{dz_{1} dz_{2}}{2} (\langle \psi(z_{1}) 
\psi(x)\rangle_{\beta} \langle \psi(z_{2}) \psi(y)\rangle_{\beta} +
(-1)^{p}\langle \psi(z_{2})
\psi(x) \rangle_{\beta} \langle \psi(z_{1}) \psi(y) 
\rangle_{\beta})\nonumber\\ 
&& \times ~\left\langle 
\frac{G[\psi]~{\stackrel{\leftarrow}{\delta^{2}}}}{\delta  
\psi(z_{1}) \delta \psi(z_{2})} F[\psi]\right\rangle_{\beta}\nonumber\\
&+& \int dz_{1} dz_{2} (\langle \psi(z_{1}) 
\psi(x)\rangle_{\beta} \langle \psi(y) \psi(z_{2})\rangle_{\beta} +
(-1)^{p} \langle \psi(x) \psi(z_{2}) 
\rangle_{\beta}\langle \psi(z_{1})
\psi(y) \rangle_{\beta})\nonumber\\
&& \times ~ \left\langle 
\frac{G[\psi]{\stackrel{\leftarrow}{\delta}}}{\delta 
\psi(z_{1})} \frac{
{\stackrel{\rightarrow}{\delta}}F[\psi]}{\delta \psi(z_{2})} 
\right\rangle_{\beta}\nonumber\\
&+& \langle \psi(x) \psi(y) \rangle_{\beta} \langle G[\psi] F[\psi] 
\rangle_{\beta}.
\tselea{S-D8}
\end{eqnarray}

\noindent We could proceed further having still higher powers of fields
and variations. However, there is a quite interesting generalization in
case when we have (anti-)time ordered operators. Let us have $F[\psi]=
{\cali{T}}(F[\psi])$, in this case

\begin{eqnarray}
\langle F[\psi] \rangle_{\beta} &=& \sum_{n} \left(\int dx\right)^{n} 
\alpha^{n}(\ldots) \langle {\cali{T}} (\prod_{i=1}^{n} \psi(x_{i}) 
\rangle_{\beta}\nonumber\\
&=& \sum_{n=1} \left( \int dx \right)^{n}  \frac{\alpha^{n} (\ldots)}{n} 
\sum_{i,j} \varepsilon_{P}\langle {\cali{T}}(\psi(x_{i}) \psi(x_{j})) 
\rangle_{\beta} 
\langle {\cali{T}}(\prod_{m \not= i,j}^{n} \psi(x_{m})) \rangle_{\beta} 
+ \alpha^{0}(\ldots)\nonumber\\
&=& \int dz_{1}dz_{2} \langle {\cali{T}}(\psi(z_{1}) 
\psi(z_{2}))\rangle_{\beta} \left\langle 
\frac{{\stackrel{\rightarrow}{\delta^{2}}} 
{\overline{F}}[\psi]}{\delta \psi(z_{2}) \delta \psi(z_{1})} 
\right\rangle_{\beta} + \langle F[0]\rangle_{\beta},
\tselea{S-D9}
\end{eqnarray}

\noindent where ${\overline{F}}[\psi]$ differs from $F[\psi]$ in the
replacement $\alpha^{n}(\ldots) \rightarrow \frac{\alpha^{n}(\ldots)}{n}$
($n$ starts from $1$ !). In comparison with (A.4)-(A.11), the
$\alpha^{0}(\ldots)$ (i.e. the pure $T=0$ contribution) does matter here. 
Note that $\alpha^{0}(\ldots)$ generally involves non-heat bath fields
with corresponding space-time integrations.  Similar extension is true if
$F[\psi] = {\cali{T}}_{c}(F[\psi])$, where ${\cali{T}}_{c}$ is the time
path ordering symbol. In that case

\begin{eqnarray}
\langle F[\psi] \rangle_{\beta}  
&=& \sum_{n} \left( \int_{C} dx \right)^{n}
\alpha^{n}(\ldots) \langle {\cali{T}}_{c} (\prod_{p=1}^{n} \psi(x_{p}))
\rangle_{\beta}\nonumber\\
&=& \int_{C} dz_{1}dz_{2} \langle {\cali{T}}_{c}(\psi(z_{1})
\psi(z_{2}))\rangle_{\beta} \left\langle 
\frac{{\stackrel{\rightarrow}{\delta^{2}}}
{\overline{F}}[\psi]}{\delta \psi(z_{2}) \delta \psi(z_{1})}
\right\rangle_{\beta}\nonumber\\
&+& \langle F[0] \rangle_{\beta},
\tselea{S-D10}
\end{eqnarray}

\vspace{2mm}

\noindent with $\int_{C} dx = \int_{C} dt \int_{V} d{\vect{x}}$ and
$\frac{\delta \psi(x)}{\delta \psi(y)}= \delta_{C}(x-y)$ \footnote{A
contour $\delta$-function $\delta_{C}(x-y)$ is defined as $\int_{C} dz
\delta_{C}(z-z^{'}) f(z) = f(z^{'})$, see {\tsecite{Mills, NS2}}.}. Wick's
theorem for the ${\cali{T}}_{C}$-oriented product of fields has an obvious
form

\begin{equation} \langle {\cali{T}}_{C} ( \psi (x_{1}) \ldots \psi 
(x_{2n}) ) 
{\rangle}_{\beta}= \sum_{\stackrel{j}{j \not= i}} \varepsilon_{P} \langle
{\cali{T}}_{C} ( \psi (x_{i})\psi (x_{j}) ) {\rangle}_{\beta} ~\langle
{\cali{T}}_{C} ( \prod_{k \not= i;j}\psi(x_{k})) {\rangle}_{\beta}.
\tseleq{wick6} \end{equation}
 
\noindent This can be directly derived from Wick's theorem (\tseref{wick2}), 
realizing that

\begin{equation}
{\cali{T}}_{C}(\psi(x_{1}) \ldots \psi(x_{m})) = \sum_{P} \varepsilon_{P} 
\theta_{C}(t_{P_{1}}, \ldots, t_{P_{m}}) \psi(x_{P_{1}}) \ldots 
\psi(x_{P_{m}}),
\end{equation}

\noindent where P refers to the permutation of the indices and 
$\theta_{C}(t_{1},\ldots , t_{m})$  being a contour step function 
\tsecite{EM} defined as

\begin{equation}
\theta_{C}(t_{1}, \ldots, t_{m})= \left\{ \begin{array}{ll}
                                    1&\mbox{($t_{1}, \ldots, t_{m}$ 
\noindent are ${\cali{T}}_{C}$-oriented along $C$)} \\
                                    0&\mbox{(otherwise)}
\end{array}
\right.
\end{equation}
 
\vspace{2mm}

\begin{figure}[h]
\vspace{4mm}
\epsfxsize=11cm
\centerline{\epsffile{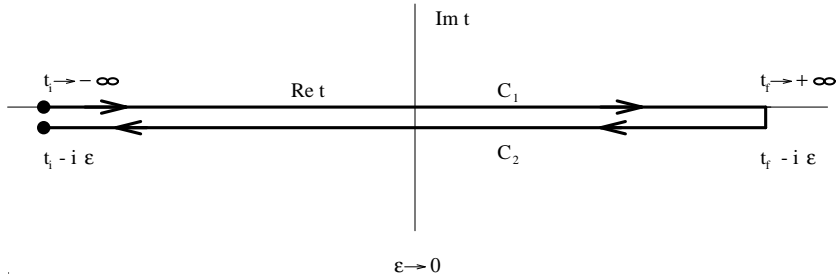}}
\caption{\em The Keldysh-Schwinger time path.}
\label{fig18}
\vspace{4mm}
\end{figure}

\noindent Particularly important is the Keldysh-Schwinger path \tsecite{LB, 
KL, EM}, see Fig.\ref{fig18}.

\noindent In the latter case 

\begin{eqnarray}
\langle F[\psi] \rangle_{\beta} &=& \int_{C_{1}} dz_{1} dz_{2} 
\langle{\cali{T}}(\psi(z_{1}) \psi(z_{2})) \rangle_{\beta} \left\langle 
\frac{{\stackrel{\rightarrow}{\delta^{2}}} {\overline{F}}[\psi]}{\delta 
\psi(z_{2}) \delta \psi(z_{1})} \right\rangle_{\beta}\nonumber\\
&+& \int_{C_{2}} dz_{1} dz_{2} 
\langle{\overline{\cali{T}}}(\psi(z_{1}) \psi(z_{2})) \rangle_{\beta} 
\left\langle
\frac{{\stackrel{\rightarrow}{\delta^{2}}} {\overline{F}}[\psi]}{\delta 
\psi(z_{2}) \delta \psi(z_{1})} \right\rangle_{\beta}\nonumber\\
&+& 2 \int_{C_{2}} dz_{1} \int_{C_{1}} dz_{2}
\langle  \psi(z_{1}) \psi(z_{2}) \rangle_{\beta}
\left\langle
\frac{{\stackrel{\rightarrow}{\delta^{2}}} {\overline{F}}[\psi]}{\delta 
\psi(z_{2}) \delta \psi(z_{1})} \right\rangle_{\beta}\nonumber\\
&+& \langle F[0] \rangle_{\beta}.
\tselea{S-D11}
\end{eqnarray}

\noindent Application to the product $G[\psi]F[\psi]$ with $F[\psi] = 
{\cali{T}}_{C_{1}}(F[\psi])$ and $G[\psi] = {\cali{T}}_{C_{2}}(G[\psi])$ is 
straightforward and reads

\begin{eqnarray} \langle G[\psi] F[\psi] \rangle_{\beta} &=&\int
dz_{1}dz_{2} \langle {\overline{\cali{T}}}(\psi(z_{1})
\psi(z_{2}))\rangle_{\beta}\left\langle ~
{\overline{\frac{G[\psi]{\stackrel{\leftarrow}{\delta^{2}}}}{\delta 
\psi(z_{2}) 
\delta \psi(z_{1})}F[\psi]}}~ \right\rangle_{\beta}\nonumber\\ &+&\int
dz_{1}dz_{2} \langle {\cali{T}}(\psi(z_{1}) \psi(z_{2}))\rangle_{\beta}
\left\langle ~ 
{\overline{G[\psi]\frac{{\stackrel{\rightarrow}{\delta^{2}}}F[\psi]}{\delta
\psi(z_{2}) \delta \psi(z_{1})}}} ~\right\rangle_{\beta}\nonumber\\ &+& 2
\int dz_{1}dz_{2} \langle \psi(z_{1}) \psi(z_{2}) \rangle_{\beta}
\left\langle ~ 
{\overline{\frac{G[\psi]{\stackrel{\leftarrow}{\delta}}}{\delta 
\psi(z_{1})}
\frac{{\stackrel{\rightarrow}{\delta}} F[\psi]}{\delta \psi(z_{2})}}} ~ 
\right\rangle_{\beta}\nonumber\\
&+& \langle G[0] F[0] \rangle_{\beta}, \tselea{S-D13} 
\end{eqnarray}

\noindent where the overlining indicates that we work with
$\frac{\alpha^{n}(\ldots) \beta^{m}(\ldots)}{n+m}$ instead of
$\alpha^{n}(\ldots) \beta^{m}(\ldots)$, we have also abbreviated
$\int_{C_{1}} dz$ to $\int dz \int_{C_{1}} dz$. We should also emphasize
that $\frac{\delta \psi(x)}{\delta \psi(y)}$ used in (\tseref{S-D13}) is
$\delta (x-y)$ rather than $\delta_{C}(x-y)$. 

\vspace{3mm}

\noindent In Eq.(\tseref{average2}) it has been used the inverted version of
(\tseref{S-D13}), namely

\begin{eqnarray} \langle (G[\psi] F[\psi])^{'} \rangle_{\beta} &=&\int
\frac{dz_{1}dz_{2}}{2} \langle {\overline{\cali{T}}}(\psi(z_{1})
\psi(z_{2}))\rangle_{\beta}\left\langle 
{\frac{G[\psi]{\stackrel{\leftarrow}{\delta^{2}}}}{\delta
\psi(z_{2}) \delta \psi(z_{1})}F[\psi]} \right\rangle_{\beta}\nonumber\\
&+&\int \frac{dz_{1}dz_{2}}{2} \langle {\cali{T}}(\psi(z_{1})
\psi(z_{2}))\rangle_{\beta} \left\langle
G[\psi]\frac{{\stackrel{\rightarrow}{\delta^{2}}}F[\psi]}{\delta 
\psi(z_{2}) \delta
\psi(z_{1})}\right\rangle_{\beta}\nonumber\\ &+& \int dz_{1}dz_{2}
\langle \psi(z_{1}) \psi(z_{2} \rangle_{\beta} \left\langle {\frac{
G[\psi]{\stackrel{\leftarrow}{\delta}}}{\delta \psi(z_{1})} 
\frac{{\stackrel{\rightarrow}{\delta}} F[\psi]}{\delta 
\psi(z_{2})}} \right\rangle_{\beta}, \tselea{S-D12} 
\end{eqnarray}

\noindent Here $(G[\psi]F[\psi])^{'}$ has the coefficients
$\alpha^{n}(\ldots)\beta^{m}(\ldots)\frac{(n+m)}{2}$ instead of
$\alpha^{n}(\ldots)\beta^{m}(\ldots)$. Note, that the
$\alpha^{0}(\ldots)\beta^{0}(\ldots)$ does not contributes and thus we do
not have any pure $T=0$ contributions. Eq.(\tseref{S-D12}) has a natural
interpretation. Whilst the LHS tells us, that from each thermal diagram
(constructed out of $\langle G[\psi] F[\psi] \rangle_{\beta}$) with
$\frac{n+m}{2}$ internal heat-bath particle lines we must take $n+m$
identical copies, the RHS says, that this is virtually because we sum over
all possible distributions of one heat-bath particle line inside of the
given diagram. The pictorial expression of (\tseref{S-D12}) is depicted in 
Fig.\ref{fig19} .

\begin{figure}[h]
\vspace{3mm}
\epsfxsize=15cm
\centerline{\epsffile{fig19.eps}}
\caption{{\it Diagrammatic equivalent of Eq.(\tseref{S-D12}). The cut 
 separates areas constructed out of $F[\psi]$ and 
$G[\psi]$.}} \label{fig19}
\vspace{3mm}
\end{figure}

\typeout{--- Extra page for bibliography ---}

\end{document}